\documentclass[%
reprint,
superscriptaddress,
amsmath,amssymb,
aps,
prapplied,
]{revtex4-2}
\usepackage[utf8]{inputenc}
\usepackage[english]{babel}
\usepackage{blindtext}
\usepackage{graphicx}   
\graphicspath{{figs/}}   
\usepackage{dcolumn}
\usepackage{bm}
\usepackage[hidelinks]{hyperref}
\usepackage{epstopdf}
\usepackage{braket}
\usepackage{upgreek}
\usepackage{amsmath}
\usepackage{siunitx}
\usepackage{changes}
\usepackage{natbib}
\usepackage[title]{appendix}

\usepackage{multirow}

\begin{document}

\title{Non-degenerate parametric amplifiers based on dispersion engineered Josephson junction arrays}

\author{Patrick Winkel}
\email{both authors contributed equally}
\affiliation{Physikalisches Institut, Karlsruhe Institute of Technology, 76131 Karlsruhe, Germany}

\author{Ivan Takmakov}
\email{both authors contributed equally}
\affiliation{Physikalisches Institut, Karlsruhe Institute of Technology, 76131 Karlsruhe, Germany}
\affiliation{Institute of Nanotechnology, Karlsruhe Institute of Technology, 76344 Eggenstein-Leopoldshafen, Germany}
\affiliation{Russian Quantum Center, National University of Science and Technology MISIS, 119049 Moscow, Russia}

\author{Dennis Rieger}
\affiliation{Physikalisches Institut, Karlsruhe Institute of Technology, 76131 Karlsruhe, Germany}

\author{Luca Planat}
\affiliation{Institut Néel, CNRS and Université Joseph Fourier, Grenoble, France}

\author{Wiebke Hasch-Guichard}
\affiliation{Institut Néel, CNRS and Université Joseph Fourier, Grenoble, France}

\author{Lukas Grünhaupt}
\affiliation{Physikalisches Institut, Karlsruhe Institute of Technology, 76131 Karlsruhe, Germany}

\author{Nataliya Maleeva}
\affiliation{Physikalisches Institut, Karlsruhe Institute of Technology, 76131 Karlsruhe, Germany}

\author{Farshad Foroughi}
\affiliation{Institut Néel, CNRS and Université Joseph Fourier, Grenoble, France}

\author{Fabio Henriques}
\affiliation{Physikalisches Institut, Karlsruhe Institute of Technology, 76131 Karlsruhe, Germany}

\author{Kiril Borisov}
\affiliation{Institute of Nanotechnology, Karlsruhe Institute of Technology, 76344 Eggenstein-Leopoldshafen, Germany}

\author{Julian Ferrero}
\affiliation{Physikalisches Institut, Karlsruhe Institute of Technology, 76131 Karlsruhe, Germany}

\author{Alexey V. Ustinov}
\affiliation{Physikalisches Institut, Karlsruhe Institute of Technology, 76131 Karlsruhe, Germany}
\affiliation{Russian Quantum Center, National University of Science and Technology MISIS, 119049 Moscow, Russia}

\author{Wolfgang Wernsdorfer}
\affiliation{Physikalisches Institut, Karlsruhe Institute of Technology, 76131 Karlsruhe, Germany}
\affiliation{Institute of Nanotechnology, Karlsruhe Institute of Technology, 76344 Eggenstein-Leopoldshafen, Germany}
\affiliation{Institut Néel, CNRS and Université Joseph Fourier, Grenoble, France}

\author{Nicolas Roch}
\email{nicolas.roch@neel.cnrs.fr}
\affiliation{Institut Néel, CNRS and Université Joseph Fourier, Grenoble, France}

\author{Ioan M. Pop}
\email{ioan.pop@kit.edu}
\affiliation{Physikalisches Institut, Karlsruhe Institute of Technology, 76131 Karlsruhe, Germany}
\affiliation{Institute of Nanotechnology, Karlsruhe Institute of Technology, 76344 Eggenstein-Leopoldshafen, Germany}

\date{\today}

\begin{abstract}
Determining the state of a qubit on a timescale much shorter than its relaxation time is an essential requirement for quantum information processing. With the aid of a non-degenerate parametric amplifier, we demonstrate the continuous detection of quantum jumps of a transmon qubit with $90\,\%$ fidelity in state discrimination. Entirely fabricated with standard two-step optical lithography techniques, this type of parametric amplifier consists of a dispersion engineered Josephson junction (JJ) array. By using long arrays, containing $10^3\,\mathrm{JJs}$, we can obtain amplification at multiple eigenmodes with frequencies below $10\,\mathrm{GHz}$, which is the typical range for qubit readout. Moreover, by introducing a moderate flux tunability of each mode, employing superconducting quantum interference device (SQUID) junctions, a single amplifier device could potentially cover the entire frequency band between 1 and 10\,GHz. 
\end{abstract}


\maketitle

\section{Introduction}
\begin{figure*}[t!]
\begin{center}
\includegraphics[width=6.67in]{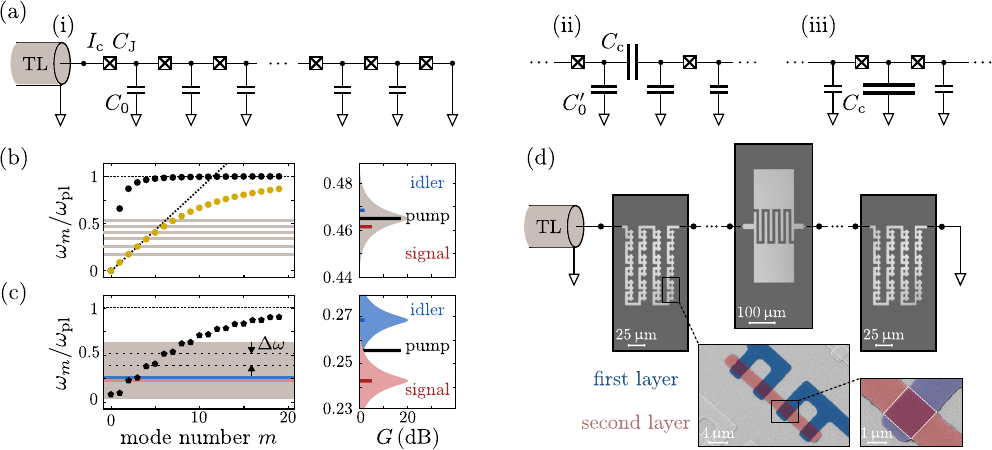}
\caption{\textbf{Dispersion engineering and optical lithography implementation of the JJ array.} \textbf{a)} Schematic circuit diagram of a Josephson Junction (JJ) array resonator. The array consists of $N$ identical dc-SQUIDs in series (for simplicity shown here as single boxes) with Josephson energy $E_\mathrm{J} = \Phi_\mathrm{0} I_\mathrm{c} / 2 \pi$ and charging energy $E_\mathrm{c} = e^2 / 2 C_\mathrm{J}$. The JJs are connected by superconducting islands with capacitance to ground $C_\mathrm{0}$. (i) The array is galvanically coupled to a $50\,\Upomega$ on-chip transmission line and terminated to ground at the other end. We engineer the dispersion relation by introducing a capacitance $C_\mathrm{c}$ in the center of the array, either (ii) in series with $C_\mathrm{J}$ or (iii) in parallel with $C_\mathrm{0}$. \textbf{b)} Calculated dispersion relation of a short ($N = 180\,\mathrm{JJs}$, black) and long ($N = 1800\,\mathrm{JJs}$, yellow) JJ array resonator for a typical value $C_\mathrm{J} / C_\mathrm{0} = 2500$. The eigenfrequencies $\omega_\mathrm{m}$ are normalized to the plasma frequency $\omega_\mathrm{pl} = \sqrt{8 E_\mathrm{J} E_\mathrm{c}}$. For the short array, only a single physical mode is in the linear regime of the dispersion relation, while the rest of the eigenmodes accumulate near $\omega_\mathrm{pl}$. By increasing the length of the array the dispersion relation flattens and several eigenmodes populate the linear regime, which is indicated by the black dotted line. \textbf{c)} Eigenmode spectrum of a long JJ array resonator ($N = 1800\,\mathrm{JJs}$) engineered according to panel (ii) or (iii). Due to the center capacitance, pairs of neighbouring modes hybridize forming a dimer each. The frequency splitting between dimer modes, $2 J_\mathrm{n}$, depends among others on the mode number $n$ and the value of $C_\mathrm{c}$, and we design it to be comparable to the mode linewidth $\kappa_\mathrm{n}$, in the range of several hundreds of MHz. As indicated in the right hand panels, each individual mode can be used for degenerate amplification, while each dimer is suitable for non-degenerate amplification. \textbf{d)} Optical microscope images of various sections of the JJ array, including the interdigitated capacitor $C_\mathrm{c}$ in the center. The array consists of optically fabricated dc-SQUIDs on a sapphire substrate arranged in a meander structure comprising 6 SQUIDs per meander. The SQUID-loop area is $A_\mathrm{L} \approx 4 \times 4 \mathrm{\upmu m^2}$. The false-colored Scanning Electron Microscopy (SEM) images in the bottom panels depict two neighbouring SQUIDs and a single JJ, respectively. In blue we highlight the first aluminum layer and in red the second. From the measured junction area $A_{\mathrm{J}} \approx 3.1\times 3.2\,\mathrm{\upmu m^2}$ we estimate a Josephson capacitance $C_\mathrm{J} \approx 500\,\mathrm{fF}$.}
\label{fig_circuit_model}
\end{center}
\end{figure*}

Low noise microwave amplifiers constitute an essential prerequisite for the implementation of fast, high fidelity quantum state detection \cite{Mallet2009,Abdo11,Vijay2011,Riste12,Lin13,Walter17} in quantum information processing with superconducting quantum bits (qubits) dispersively coupled to readout resonators \cite{Wallraff04,Blais04}. Although in principle the strength of the readout signal can be increased well above the noise of commercial high electron mobility transistor (HEMT) amplifiers, this typically results in an increase of the qubit's energy relaxation rate \cite{Lescanne18, Verney19}, which overall degrades the readout fidelity. Over the last decade, this limitation in signal-to-noise ratio (SNR) has been successfully overcome thanks to the development of superconducting parametric amplifiers \cite{Yurke06, Castellanos-Beltran2007, Yamamoto2008,Bergeal2010,Hatridge11,Roch12, Mutus13, Eichler14_DPA} which add less noise, down to the quantum limit \cite{Caves1982}. \\
\\
In superconducting parametric amplifiers, the non-linearity required to transfer energy from a strong classical pump tone to a weak quantum signal \cite{Franken61, Clerk2010} is provided by low-loss inductive elements, namely Josephson junctions (JJ) \cite{Jos62,Likharev86,Yurke89}, or thin films of disordered superconductors with intrinsically high kinetic inductance \cite{Chin92,Tholen09,Maleeva18}. These non-linear elements are either embedded in a resonant tank circuit \cite{ROY16} or in a dispersion engineered microwave transmission line \cite{Eom12,Bockstiegel2014,Macklin15,Zorin16,PlanatTWPA19}. In the first case, amplification only occurs in the vicinity of the standing-wave eigenfrequency of the circuit, while in the case of travelling wave parametric amplifiers, the applied tones interact during propagation along the transmission line in a much larger frequency band, covering several GHz.\\
\\ 
Ideally, a parametric amplifier offers a signal power gain $G \geq 20\,\mathrm{dB}$ in a frequency band larger than the linewidth of the readout resonator, a saturation power well above the single photon regime \cite{Eichler14,Zhou14,Fedorov15,Boutin17,Liu17,Frattini18,Planat19}, and  isolation of the qubit-resonator system from the strong pump \cite{Abdo2013,Sliwa15,Lecocq17}. Although there has been impressive progress in the development of broadband travelling wave parametric amplifiers \cite{Eom12,Bockstiegel2014,Macklin15,Zorin16,PlanatTWPA19}, optimizing these figures of merit for a specific application is generally simpler for a standing-wave parametric amplifier. \\
\\
We present a standing-wave parametric amplifier design based on SQUID-arrays containing up to 1800 SQUIDs, with an engineered dispersion relation realizing pairs of hybridized modes (dimers) suitable for non-degenerate parametric amplification [cf. Fig\,\ref{fig_circuit_model}]. By applying a strong, single-frequency pump tone in-between the hybridized modes, we demonstrate signal power gains exceeding $20\,\mathrm{dB}$ for up to four dimers in a single device [cf. Fig\,\ref{fig_gain_measurements}], over an instantaneous bandwidth between $5$ and $15\,\mathrm{MHz}$. The pump tone frequency is detuned from the signal by hundreds of MHz, which enables its filtering. We refer to these devices as Dimer Josephson Junction Array Amplifiers (DJJAAs). We note that a similar idea was developed in parallel in Ref.\,\cite{SivJAMPA19}, where several modes of SNAIL \cite{Frattini17} (superconducting nonlinear asymmetric inductive element) arrays were used for amplification.\\
\\
This article is organized as follows: In Sec.\,\ref{SEC:Concept}, we present the DJJAA concept, followed by the effective circuit model in Sec.\,\ref{SEC:CircuitModel}, which is used to calculate the DJJAA dispersion relation, eigenfunctions and first order nonlinear coefficients. In Sec.\,\ref{SEC:devicefabrication} we describe the optical fabrication process. In Sec.\,\ref{SEC:Gainmeasurements} we show power gain measurements for three DJJAAs. For the longest investigated array (1800 SQUIDs), we observe non-degenerate amplification exceeding $20\,\mathrm{dB}$ for four individual dimers in a single DJJAA device, epitomizing the potential of this amplifier design. Section\,\ref{SEC:noise_characterization} is devoted to noise characterization. Using an array of 1200 SQUIDs, we calibrate the measurement efficiency $\eta$ of our setup \cite{Hatridge13} by observing quantum jumps of a transmon qubit \cite{Koch07}. We find $\eta \approx 0.13$ for the whole setup and $\eta_\mathrm{DJJAA} \ge 0.29$ for our parametric amplifier, which is comparable to values reported in literature \cite{Hatridge13,Vool14,Macklin15}. In Sec.\,\ref{SEC:conclusion} we conclude by summarizing the main results. 
\section{Concept}
\label{SEC:Concept}
The eigenmode spectrum of a Josphson junction array (JJA) [cf. Fig.\,\ref{fig_circuit_model}a)] contains $N$ eigenmodes, given by the total number of JJs \cite{Fazio01,Masluk12,Weissl15,Krupko18,Muppalla18,Planat19}. For small frequencies, the effective wavelength of the eigenmodes is much larger than the distance between neighbouring JJs, and the mode frequency increases almost linearly with the mode index $m$ [cf. Fig.\,\ref{fig_circuit_model}b)]. The slope of this linear regime is determined by the square root of the ratio between the capacitance per unit length arising from the Josephson capacitance $C_\mathrm{J}$, and the capacitance to ground $C_0$ \cite{Hutter08}. When the effective wavelength of the eigenmodes becomes comparable to the distance between the JJs, the dispersion relation becomes non-linear, eventually saturating at the self-resonance frequency of a single JJ, denoted plasma frequency $\omega_\mathrm{pl} \approx 1 /\sqrt{L_\mathrm{J} C_\mathrm{J}}$. Here, $L_\mathrm{J}$ is the Josephson inductance, $C_\mathrm{J}$ is the Josephson capacitance, and we neglect the contribution of $C_0$. Due to the non-linearity of the JJ cosine potential, each eigenmode is itself non-linear in power. As we will show in the following, this non-linearity can be used for parametric amplification. \\
\\
In general, parametric amplifiers are classified into degenerate and non-degenerate designs, depending on whether the signal ($\omega_\mathrm{s}$) and the idler tones ($\omega_\mathrm{i}$) occupy the same or different physical modes \cite{ROY16}. In the latter case, protecting the quantum circuit under investigation from the influence of the strong pump tone becomes considerably simpler, since the signal and pump tone are detuned in frequency. \\
In order to obtain non-degenerate amplification, we introduce a capacitor in the center of the JJ array, which depending on the design [cf. Fig.\,\ref{fig_circuit_model}a) panel (ii) and (iii)], it either splits the array in two capacitively coupled sections, or it capacitively shunts the central island to the ground. In either of these cases, the capacitor breaks the symmetry between even and odd modes, and creates pairs of hybridized modes [cf. Fig.\,\ref{fig_circuit_model}c)]. Each pair, denoted dimer in the following, is suitable for non-degenerate parametric amplification by applying a pump tone in-between the two modes. In this four-wave-mixing process, two pump photons are converted into a signal and idler photon ($\omega_\mathrm{s} + \omega_\mathrm{i} = 2 \omega_\mathrm{p}$), similar to the scheme presented in Ref.\,\cite{Eichler14_DPA}. The device works in reflection, with the pump added to the signal. \\ 
\\
The intrinsic limitation in instantaneous bandwidth for standing wave parametric amplifiers is overcome by employing SQUID junctions with flux-tunable critical current $I_\mathrm{c} \left( \Phi \right)$. Tuning the device frequency by lowering $I_\mathrm{c}$ also increases the impact of higher order nonlinear terms arising from the Josephson potential, which will eventually limit the amplifier performance in terms of dynamic range \cite{Fedorov15,Boutin17,Liu17} and therefore bound the tuning bandwidth. In order to mitigate the effect of higher order terms and maximize the saturation power of the device, we use long arrays of JJs similar to the approaches in Refs.\,\cite{Yurke96,Castellanos-Beltran2007,Lahteenmaki13,Eichler14,Zhou14,Planat19}.\\
\\
The tunable bandwidth of the DJJAA is given by the flux tunability of each dimer suitable for amplification. Since the frequency difference $\Delta \omega$ between neighbouring dimers decreases as the number of SQUIDs is increased [cf. Fig.\,\ref{fig_circuit_model}c)], we can imagine that $\Delta \omega$ can be reduced to values comparable to the flux tunable bandwidth of each dimer. In this case, the effective tunable bandwidth of the DJJAA would span over the entire linear part of the dispersion relation, which is typically several GHz wide (highlighted in grey in Fig.\,\ref{fig_circuit_model}c)). As discussed in Sec.\,\ref{SEC:Gainmeasurements}, we demonstrate a step in this direction, by showing power gain reaching $20\,\mathrm{dB}$ for four different dimers in the same device, spread over a frequency range of $4\,\mathrm{GHz}$. 
\section{Circuit Model}
\label{SEC:CircuitModel}
In order to calculate the dispersion relation of our DJJAAs, we derive the system Lagrangian from an effective circuit model. The SQUID arrays consist of $N$ SQUIDs in series, with zero-field critical current $I_\mathrm{c}$ and Josephson capacitance $C_\mathrm{J}$, which are connected by superconducting islands with capacitance $C_0$ to ground [cf. panel (i) in Fig.\,\ref{fig_circuit_model}a)]. For simplicity, additional capacitances arising between islands due to long-term Coulomb interactions mediated by the shared ground plane \cite{Krupko18} are neglected in our model. The input port of the array is galvanically coupled to a $50\,\Omega$ on-chip transmission line, while the other end is terminated to ground. In the center of the array we introduce an additional capacitance $C_\mathrm{c}$ either in series with $C_\mathrm{J}$ [cf. panel (ii) in Fig.\,\ref{fig_circuit_model}a)] or in parallel to $C_\mathrm{0}$ [cf. panel (iii) in Fig.\,\ref{fig_circuit_model}a)].\\
\\
Although both options for the dispersion engineering of the array [cf. panel (ii) and panel (iii) in Fig.\,\ref{fig_circuit_model}a)] result in dimers which can be used for non-degenerate amplification, in the following, we will only discuss the first approach in detail. \\
\\
Due to the physical dimension of the center capacitor plates, the capacitance $C_\mathrm{0}^\prime$ to ground on the central nodes with indices $N/2$ and $N/2 +1$ is enhanced ($C_\mathrm{0}^\prime \gg C_\mathrm{0}$) in the first case. The Lagrangian of our system writes
\begin{equation}
\begin{split}
\mathcal{L} = &\sum_{i = 1}^{N/2-1} \frac{C_\mathrm{0}}{2} \dot{\Phi}_{i}^2 + \sum_{i = \frac{N}{2} + 2}^{N} \frac{C_\mathrm{0}}{2} \dot{\Phi}_{i}^2 \\
& + \frac{C_\mathrm{0}^\prime}{2} \left( \dot{\Phi}_{N/2}^2 + \dot{\Phi}_{N/2+1}^2 \right) + \frac{C_\mathrm{c}}{2} \left( \dot{\Phi}_{N/2 + 1} - \dot{\Phi}_{N/2}\right)^2 \\
& + \sum_{i = 0}^{N/2 - 1} \frac{C_\mathrm{J}}{2} \left( \dot{\Phi}_{i+1} - \dot{\Phi}_{i} \right)^2 + \sum_{i = N/2 + 1}^{N} \frac{C_\mathrm{J}}{2} \left( \dot{\Phi}_{i+1} - \dot{\Phi}_{i} \right)^2 \\
& - \sum_{i = 0}^{N/2 - 1} E_\mathrm{J} \cos(\phi_{i+1} - \phi_{i}) - \sum_{i = N/2 + 1}^{N} E_\mathrm{J} \cos(\phi_{i+1} - \phi_{i}). 
\end{split}
\label{LAG:ARRAYC}
\end{equation}
Here, $E_\mathrm{J} = \Phi_0 I_\mathrm{c} / 2 \pi$ is the Josephson energy and $\phi_{n}$ is the superconducting phase of the $n$-th island, with the corresponding node flux $\Phi_{n} = \Phi_0 \phi_{n} / 2 \pi$. For a system galvanically coupled to the environment on both ends, the boundary conditions are $\Phi_{0} = \Phi_{N+1} = 0$. \\
In the limit of small circulating current $I \ll I_\mathrm{c}$, the phase drop $\phi_{{i}+1} - \phi_{i}$ across each JJ is small, and we can describe the JJs as linear inductors with kinetic inductance $L_\mathrm{J} = \Phi_0 / 2 \pi I_\mathrm{c}$. By introducing the node flux vector $\vec{\Phi} = (\Phi_0, .., \Phi_{N+1})$, Eq.\,\ref{LAG:ARRAYC} can be rewritten in matrix representation
\begin{equation}
\mathcal{L} = \frac{1}{2} \dot{\vec{\Phi}}^\mathrm{T} \tilde{C} \dot{\vec{\Phi}} - \frac{1}{2} \vec{\Phi}^\mathrm{T} \tilde{L}^{-1} \vec{\Phi},
\label{LAG:flux}
\end{equation}
where $\tilde{C}$ and $\tilde{L}$ are the capacitance and inductance matrices, respectively [cf. App.\,\ref{A_inductance_matrix}]. Following Eq.\,\ref{LAG:flux}, the eigenfrequencies $\omega_{m}$ of the system are calculated by numercially solving the eigenvalue problem \cite{Weissl15}
\begin{equation}
\tilde{C}^{- 1/2} \tilde{L}^{-1} \tilde{C}^{- 1/2} \vec{\Psi}_{m} = \omega_{m}^2 \vec{\Psi}_{m},
\label{EQ:EIGENPROBLEM}
\end{equation}
where the corresponding eigenvectors $\vec{\Psi}_{m}$ are related to the node flux eigenvectors $\vec{\Phi}_{m}$, which carry information about the modes' spatial distribution along the array \cite{Masluk12}.\\
\\
In order to illustrate the effect of increasing the number of SQUIDs $N$ in the array, in Fig.\,\ref{fig_circuit_model}b) we plot the dispersion relation obtained from Eq.\,\ref{EQ:EIGENPROBLEM} for two arrays with  $N = 180$ (black) and  $N = 1800$ (yellow), while maintaining a fixed ratio $C_\mathrm{J} / C_\mathrm{0} = 2500$. For clarity, the eigenfrequencies $\omega_{m}$ are normalized to the plasma frequency $\omega_\mathrm{pl}$. The longer the chain, the more eigenmodes fall into the linear regime, with decreasing frequency detuning $\Delta \omega \,(= \omega_{i+1} - \omega_{i})$ between neighbouring modes. For the long array (yellow) and assuming a typical plasma frequency $\omega_{pl} \approx 20\,\mathrm{GHz}$, the eigenmodes falling into the technologically favored frequency range below 10\,GHz are highlighted by horizontal, grey lines. As indicated by the panel on the right, each eigenmode is suitable for degenerate amplification by applying a pump tone on resonance. \\
\\
By introducing the capacitance $C_\mathrm{c}$ in the center of the array according to Fig.\,\ref{fig_circuit_model}a) panel (ii), the system exhibits symmetric and anti-symmetric pairs of hybridized modes [cf. Fig.\,\ref{fig_circuit_model}c)], denoted dimers. As indicated by the panel on the right, each dimer is suitable for non-degenerate amplification by applying a single-frequency pump tone through the input-port, in-between the two dimer modes, similar to the concept presented in Ref.\,\cite{Eichler14_DPA}. Due to this off-resonant pumping scheme, the signal frequency $\omega_\mathrm{s}$ is well detuned from the pump frequency $\omega_\mathrm{p}$, reducing pump leakage \cite{Slichter11,Hatridge13,Lescanne18}. \\
\\
In order to couple the pump tone to both dimer modes, the frequency difference $2 J_{n}$ between the two modes of a dimer, where $n$ denotes the dimer index, is designed to be comparable to the amplifiers' linewidth. In our case $J_n$ is on the order of several hundreds of MHz and it depends on the size of the center capacitance $C_\mathrm{c}$ and its energy participation ratio in each mode, which is related to the total length of the chain. Therefore, as a rule of thumb, the values of $C_\mathrm{c}$ and $N$ are linearly related. From calculations based on Eq.\,\ref{EQ:EIGENPROBLEM}, we choose $C_\mathrm{c} = 45\,\mathrm{fF}$ and $C_\mathrm{c} = 30\,\mathrm{fF}$ for $N = 1800$ and $N = 1200$, respectively. In both cases, using finite element simulations of the capacitor geometry [cf. Fig.\,\ref{fig_circuit_model}d)], we extract a parasitic capacitance to ground $C_0^\prime = 33\,\mathrm{fF}$. \\
\\
Starting with the linear circuit model of Eq.\,\ref{LAG:flux}, we can perturbatively introduce the non-linearity arising from the Josephson potential, by expanding up to the quartic term $\propto \Delta \phi^4$ and applying the rotating wave approximation (RWA) \cite{Weissl15,Krupko18}. We obtain the Hamiltonian
\begin{equation}
\begin{split}
\textbf{H} = \sum_{m = 0}^{N-1} \hbar \omega_{m} \textbf{a}^\dagger_{m} \textbf{a}_{m} - &\sum_{m = 0}^{N-1} \frac{\hbar}{2} K_{m,m}^{} \textbf{a}^\dagger_{m} \textbf{a}_{m} \textbf{a}^\dagger_{m} \textbf{a}_{m} \\
& - \sum_{m,k = 0}^{N-1} \frac{\hbar}{2} K_{m,k} \textbf{a}^\dagger_{m} \textbf{a}_{m} \textbf{a}^\dagger_{k} \textbf{a}_{k},
\end{split}
\label{HAM:KERR_SQ}
\end{equation}
where $\mathbf{a}^\dagger_m$ and $\mathbf{a}_m$ are the bosonic single-mode field amplitude creation and annihilation operators, while $K_{m,m}$ and $K_{m,k}$ are the self-Kerr and cross-Kerr coefficients, respectively. The first term describes the harmonic system, the second term relates the frequency of the $m$-th mode $\omega_{m}$ to the mean circulating photon number $\bar{n}_{m}$ of that same mode, and the third term describes the interaction between two modes with indices $m$ and $k$. Using the same notations as in Ref.\,\cite{Weissl15}, the Kerr coefficients expressed in terms of the circuit parameters are
\begin{equation}
\begin{split}
K_{m,m} &=  \frac{2 \hbar \pi^4 E_\mathrm{J} \eta_{mmmm}}{\Phi_0^4 C_\mathrm{J}^2 \omega_{m}^2},\\
K_{m,k} &= \frac{4 \hbar \pi^4 E_\mathrm{J} \eta_{mmkk}}{\Phi_0^4 C_\mathrm{J}^2 \omega_{m} \omega_{k}}.
\end{split}
\label{EQ:KERRs}
\end{equation} 
Besides the JJ parameters $E_\mathrm{J}$ and $C_\mathrm{J}$, the Kerr coefficients depend on the eigenfunctions $\vec{\Phi}_{m}$, which give the dimensionless factors $\eta_{mmkk}$ [cf. App.\,\ref{A_Kerr}]. It is these terms which can give rise to parametric amplification under microwave pumping.\\
\\
\begin{figure*}[!t]
\begin{center}
\includegraphics[width=6.67in]{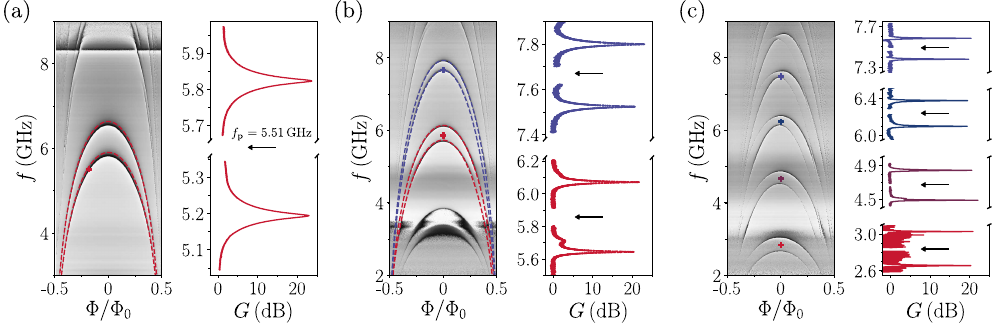}
\caption{\textbf{Phase response of the complex reflection coefficient $\arg(S_\mathrm{11})$ measured as a function of externally applied flux $\Phi$ for three dispersion engineered JJ arrays:} \textbf{a)} $N = 1200$ JJs, $I_\mathrm{c} \approx 6.1 \, \mathrm{\upmu A}$, $C_\mathrm{c} = 30\,\mathrm{fF}$, \textbf{b)} $N = 1600$ JJs, $I_\mathrm{c} \approx 3.0 \, \mathrm{\upmu A}$, $C_\mathrm{c} = 40\,\mathrm{fF}$, \textbf{c)} $N = 1800$ JJs, $I_\mathrm{c} \approx 2.3 \, \mathrm{\upmu A}$, $C_\mathrm{c} = 45\,\mathrm{fF}$. The dispersion relation is dimerized by introducing $C_\mathrm{c}$ in series with $C_\mathrm{J}$ in the center of the array [cf. panel (ii) in Fig.\,\ref{fig_circuit_model}a)].  For all plots the color scale covers the entire range from $-\pi$ (black) to $\pi$ (white). In a) and b) we show in dashed lines a typical example of numerical fits to the eigenmodes, used to calibrate the magnetic coil current and to extract the device parameters. As expected, the number of modes within a given frequency range (in our case $4-8\,\mathrm{GHz}$) increases with increasing $N$ and decreasing $I_\mathrm{c}$. By applying a strong pump tone in-between two hybridized modes, non-degenerate power gain exceeding $G_\mathrm{0} = 20\,\mathrm{dB}$ is observed for up to four pairs of modes in a single device, as shown in panel c). The arrow and cross symbols in each panel indicate the external flux bias and pump frequency, color-coded for each dimer. The horizontal features visible in the vicinity of $4$ and $8\,\mathrm{GHz}$ correspond to the frequency band of the circulator attached to the DJJAA input port.} 
\label{fig_gain_measurements}
\end{center}
\end{figure*} 
\section{Device fabrication and measurement setup}
\label{SEC:devicefabrication}
The SQUID arrays are implemented in a microstrip geometry with platinum backside metallization of thickness $t = 300\,\mathrm{nm}$, and are aranged in a meander structure comprising 6 SQUIDs per meander. The minimum feature size is chosen to be relatively large, around $3\,\si{\micro\metre}$, in order to facilitate microfabrication using optical lithography techniques for the entire device. The circuit is patternd in two separate steps, each followed by a zero angle aluminum (Al) thin film evaporation of thickness 30\,nm and 40\,nm. Before depositing the second Al layer, we apply an in-situ argon-milling cleaning step to remove the native oxide from the surface of the first Al layer \cite{Gruenhaupt17, Wu17}. The Al/AlO\textsubscript{x}/Al JJs are formed by the overlap areas between the first and second Al layer, with an area $A_\mathrm{JJ} \sim 9 - 11\,\si{\micro\metre}^2$. AlO\textsubscript{x} denotes non-stoichiometric insulating aluminum oxide grown under static oxidation in an oxygen pressure of $10\,\mathrm{mbar}$ for 2 - 4 minutes. Due to the relatively small SQUID loop area $A_\mathrm{L} \approx 4 \times 4\,\si{\micro\metre}^2$ and wire width $w \approx 3\,-\,4\, \si{\micro\metre}$ in our design, the inductive contribution of the loop $L_\mathrm{loop}$ on the flux modulation is neglectable ($L_\mathrm{loop} \ll L_\mathrm{J}$). In addition, each island adds a stray inductance $L_\mathrm{stray} \approx 20\,-\,30\,\mathrm{pH}$. \\
\\
The sapphire chips hosting the DJJAAs are glued into a copper sample holder, which has a dedicated superconducting flux coil integrated into the lid [cf. App.\,\ref{A_sample_holder}]. The on-chip transmission line is connected to the coaxial input port of the sample holder with aluminum microbonds. A microwave circulator is connected directly to the sample holder port, and is used to separate the incident from the reflected outgoing signals. The DJJAAs are anchored to the millikelvin stage of a dilution refrigerator with a base temperature $T_\mathrm{base} \approx 20 - 30\,\mathrm{mK}$, and probed in a single port reflection measurement. After reflection from the DJJAA, the outgoing signal is further amplified by two commercial amplifiers, a HEMT amplifier at $4\,\mathrm{K}$ and a room temperature amplifier.
\begin{figure*}[!t]
\begin{center}
\includegraphics[width=6.69in]{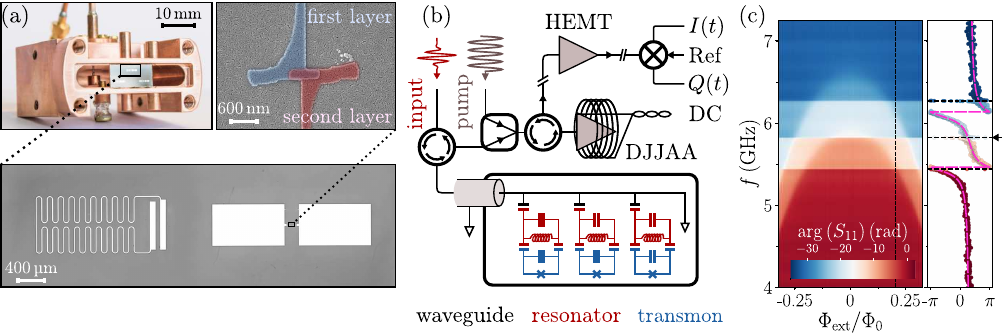}
\caption{\textbf{Transmon qubit samples.} \textbf{a)} Photograph of the copper waveguide sample holder. In the center of the waveguide, where the electric field is maximum, we place a sapphire chip with three transmon qubits, each capacitively coupled to a dedicated lumped element readout resonator. The lower panel shows an optical image of one of the transmon qubits and its readout resonator, with an estimated coupling strength $g / 2 \pi \approx 35 \, \mathrm{MHz}$ [cf. App.\,\ref{A_Qubit_chara}]. The top right panel depicts an SEM image of a single JJ. All structures are patterned using electron beam lithography, and they are deposited by shadow-angle evaporation. \textbf{b)} Circuit diagram of the measurement setup. The transmon qubits (blue) are dispersively coupled to the readout resonators (red) and mounted inside the waveguide (black). The readout signal (red arrow) is preamplified by a DJJAA and routed to a commercial High Electron Mobility Transistor amplifier (HEMT) mounted at $4\,\mathrm{K}$. At room temperature the signal is decomposed into its in-phase and out-of-phase quadratures using a heterodyne microwave interferometer. The DJJAA pump tone (grey arrow) is fed into the signal path through a commercial power combiner. The flux bias for the DJJAA is supplied by an external magnetic field coil [cf. App.\,\ref{A_sample_holder}]. \textbf{c)} Unwrapped phase of the complex reflection coefficient, $\arg(S_\mathrm{11})$, as a function of frequency $f$ and applied bias flux $\Phi_\mathrm{ext}$ measured from the input port. The three sharp, flux independent phase-rolls are due to the readout resonators, and the two broad, flux dependent features are given by an amplifier dimer [cf. Fig.\,\ref{fig_gain_measurements}a)]. The right hand panel depicts the wrapped phase response along the black dashed line for the DJJAA flux bias used to measure quantum jumps of the transmon coupled to the resonator at $f_\mathrm{r} = 5.8224\,\mathrm{GHz}$ (indicated by the black arrow). The pink line indicates a fit to the DJJAA linear response (for details see App.\,\ref{A_Dimer})}
\label{fig_Transmon_sample}
\end{center}
\end{figure*}
\section{Gain measurements}
\label{SEC:Gainmeasurements}
The left hand panels in Fig.\,\ref{fig_gain_measurements}a), b), and c) show the phase of the complex reflection coefficient $\arg(S_{11})$ as a function of probe frequency $f$ and external flux bias $\Phi$ for three different samples with a total number of $N = 1200$, $1600$ and $1800$ SQUIDs. The respective critical currents per SQUID are $I_\mathrm{c} = 6.1\,\si{\micro\ampere}, 3.0\,\si{\micro\ampere}$ and $2.3\,\si{\micro\ampere}$, with corresponding Josephson inductance $L_\mathrm{J} = 54\,\mathrm{pH},\,\,109\,\mathrm{pH}$ and $142\,\mathrm{pH}$, respectively. Due to the sweep in design parameters, we observe a single dimer in (a), three dimers in (b) and four dimers in (c). With increasing external flux, the frequency of the dimers decreases, as expected from increasing the SQUID inductance, with several higher modes becoming visible close to full SQUID frustration ($|\Phi / \Phi_0| \approx 0.5$). The dashed lines depicted for the first two samples indicate numerical fits to estimate the stray inductance $L_\mathrm{stray}$ of the circuit design and to calibrate the bias current of the superconducting field coils [cf. Appendix Sec.\,\ref{A_FluxMod}]. Furthermore, from these fits we conclude that the SQUID asymmetry ($I_\mathrm{c,1} \neq I_\mathrm{c,2}$) and the loop inductance $L_\mathrm{loop}$ can be neglected. \\
\\
The right hand panels in Fig.\,\ref{fig_gain_measurements}a), b) and c) depict the power gain $G$ in dB as a function of probe frequency $f$ when an additional pump tone of power $P_\mathrm{p}$ and frequency $f_\mathrm{p}$ is applied in-between two dimerized modes. The black arrows and cross symbols, which are color-coded individually for each dimer, indicate the external bias flux $\Phi$ and pump frequency used in each  experiment, respectively. For all dimers, the pump frequency is off-centered with respect to the low probe power response, since the mode population $\bar{n}_m$ caused by the strong pump tone shifts the dimer modes in frequency by an amount $K_{m,m} \bar{n}_m$. The observed power gain profile in the high gain limit ($G \gg 1$) is composed of two overlapping Lorentzian curves, symmetrically emerging below and above the pump tone frequency. Due to the self-Kerr coefficients $K_{m,m}$ [cf. Eq.\,\ref{EQ:KERRs}], the frequency detuning between the two maxima depends on the pump power and the pump frequency, similar to Ref.\,\cite{Eichler14_DPA}. The power gain exceeds $20\,\mathrm{dB}$, which is a typical value required to saturate the classical noise added by higher temperature amplifier stages with amplified quantum noise. \\
\\
As in the case of other Josephson parametric amplifiers (JPA), there is a compromise between the maximum of the gain, $G_0$, and the amplifier's instantaneous bandwidth $B$, defined as the full width at half maximum (FWHM), which is reflected by a constant gain-bandwidth product $\sqrt{G_0} B$ \cite{Clerk2010}. For the sample shown in Fig.\,\ref{fig_gain_measurements}a), which has the largest coupling, we find $\sqrt{G_0} B \approx 170\,\mathrm{MHz}$ at the flux-sweet spot ($\Phi = 0$). This value is in good agreement with the average of the two measured resonator linewidths $\bar{\kappa} / 2 \pi = 172\,\mathrm{MHz}$, as expected for two parametrically coupled modes \cite{Clerk2010, Eichler14_DPA}. Thanks to the SQUID junctions the frequency at which we obtain gain is flux tunable. We typically measure a tunable bandwidth of around $1\,\mathrm{GHz}$ per dimer (see App.\,\ref{A_flux_tunability}). \\
\\
The input saturation power $P_\mathrm{1dB}$, conventionally defined as the signal probe power at which the maximal power gain $G_0$ decreases by $1\,\mathrm{dB}$ ($1\,\mathrm{dB}$-compression point), reportedly scales with the ratio of amplifier linewidth and self-Kerr coefficient $\kappa_{n} / |K_{n,n}|$ \cite{Boutin17,Liu17}. In our case, this ratio depends on the dimer mode number $n$, in accordance to the corresponding spatial mode distribution $\vec{\Phi}_{n}$ [cf. Sec.\,\ref{SEC:CircuitModel}]. Thanks to the large number of SQUIDs in our design, the self-Kerr coefficients are strongly reduced for the lowest modes of the dispersion relation compared to designs with only a single or a few SQUIDs \cite{Mutus13, Hatridge11}. For this reason, $P_\mathrm{1dB}$ is found to be enhanced by an order of magnitude in SQUID array based amplifiers similar to our approach \cite{Eichler14_DPA,Planat19}. As we will decribe in the following section, this allows us to operate the DJJAA at a signal strength of $P_\mathrm{s} \geq -118\,\mathrm{dBm}$ ($\geq 420\,\mathrm{photons} \times \si{\micro\second}^{-1}$) without observing the onset of saturation. 

\section{Noise characterization}
\label{SEC:noise_characterization}
\subsection{Power calibration}
\label{SEC:power_calibration}
In order to evaluate the noise performance of a typical DJJAA parametric amplifier, we calibrate the signal power referred to its input port with a transmon qubit \cite{Koch07}, which is dispersively coupled to a dedicated readout resonator, as shown in Fig.\,\ref{fig_Transmon_sample}. In close vicinity to the resonance frequency $f_\mathrm{r}$ of the readout resonator, the signal strength can be expressed in number of measurement photons $n_\mathrm{meas}$ \cite{Hatridge13, Vool14}. In a continuous single-port reflection measurement, the measurement photon number is
\begin{equation}
n_\mathrm{meas} = \bar{n}_\mathrm{r} \frac{(\kappa_\mathrm{r} + \gamma_\mathrm{r})^2}{4 \kappa_\mathrm{r}} T_\mathrm{m} \underset{\gamma_\mathrm{r} \rightarrow 0}{\approx} \bar{n}_\mathrm{r} \frac{\kappa_\mathrm{r}}{4} T_\mathrm{m},
\label{EQ:nmeas}
\end{equation}
where $T_\mathrm{m}$ is the measurement integration time, $\bar{n}_\mathrm{r}$ is the mean number of photons circulating inside the readout resonator, and $\gamma_\mathrm{r}$ and $\kappa_\mathrm{r}$ are the resonator's internal and external decay rates, respectively. \\
\\
The qubit sample design contains three transmon qubits, each containing a single Josephson junction shunted by an in-plane plate capacitor with rectangular pads, which are capacitively coupled to a dedicated lumped-element readout resonator [cf. Fig.\,\ref{fig_Transmon_sample}a)]. The fabrication is based on the bridge free technique \cite{Lecocq11}, a shadow-angle evaporation technique, and the evaporation of two aluminum thin films on a double polished sapphire substrate, separated by an oxygen barrier. \\
\\
Each chip is mounted in a 3D-waveguide sample holder \cite{Kou18}[cf. Fig.\,\ref{fig_Transmon_sample}a)] and measured in reflection [cf. Fig.\,\ref{fig_Transmon_sample}b)]. The reflected readout signal is routed to a DJJAA using a cryogenic circulator. The pump tone, which serves as the power supply for our parametric amplifier, is feed into the signal path with a commercial power combiner. The incident and reflected (amplified) signals are again separated employing a cryogenic circulator directly mounted at the input of the DJJAA, and subsequently further amplified by two amplifier stages, a commercial HEMT amplifer with a specified noise temperature $T_\mathrm{N, HEMT} \approx 2.3\,\mathrm{K}$ mounted at $4\,\mathrm{K}$ and a room temperature amplifier with $T_\mathrm{N, RT} \approx 170\,\mathrm{K}$. \\
\\
Figure\,\ref{fig_Transmon_sample} (c) depicts the unwrapped phase of the measured complex reflection coefficient $\arg(S_\mathrm{11})$ in radians as a function of probe frequency $f$ and external flux $\Phi_\mathrm{ext}$. The frequencies of the three readout resonators, corresponding to the three qubits, are independent of the applied flux bias and are visible as sharp horizontal lines. Since the amplifier modes are much stronger coupled to the input port compared to the readout resonators ($\kappa \gg \kappa_\mathrm{r}$), the dimer modes appear as broad features. The right hand panel depicts the wrapped phase response for the flux bias indicated by the black dashed line in the left hand panel. For clarity, the readout resonator frequencies are indicated by horizontal black lines. The dashed pink line indicates the fitted phase response given by two DJJAA modes separated by $2 J / 2 \pi = 670\,\mathrm{MHz}$ and coupled to the input port with coupling rate $\kappa_+ / 2 \pi = 148\,\mathrm{MHz}$ and $\kappa_- / 2\pi = 139\,\mathrm{MHz}$ (see App.\,\ref{A_Dimer} for details).\\
\\
The frequency of the readout resonator ($f_\mathrm{r} = 5.8224\,\mathrm{GHz}$) coupled to the transmon qubit used to calibrate the DJJAA noise [cf. Sec.\,\ref{SEC:noise_characterization} (C)] is indicated by the small black arrow on the right hand side. 
\begin{figure}[!t]
\begin{center}
\includegraphics[width = 1\columnwidth]{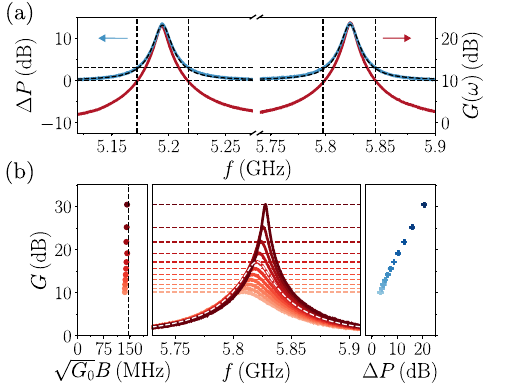}
\caption{\textbf{Noise visibility.} \textbf{a)} Noise visibility $\Delta P$ (left axis) and power gain $G$ (right axis) in dB as a function of frequency for the flux bias and pump power used during the qubit experiments. The maximum power gain $G_\mathrm{0} = 23.2\,\mathrm{dB}$ and corresponding bandwidth (FWHM) $B \approx 9.2\,\mathrm{MHz}$ are extracted from Lorentzian fits. The observed noise visibility at maximum gain is $\Delta P = 14.2\,\mathrm{dB}$. \textbf{b) } Power gain $G$ in dB as a function of frequency for various pump powers. For each curve $G_\mathrm{0}$ and $B$ are extracted from Lorentzian fits to the amplifier response, generically indicated by the dashed white line (central panel). The calculated gain-bandwidth product $\sqrt{G_\mathrm{0}} B \approx 143\,\mathrm{MHz}$ is in good agreement with the theoretical prediction (black line, left panel). For each pump condition the noise visibility $\Delta P$ was measured, too, with the visibility at maximum power gain plotted in blue.}
\label{fig_noise_visibility}
\end{center}
\end{figure}
\subsection{Noise visibility}
During the qubit experiments, we operate our parametric amplifier at a power gain $G_\mathrm{0} = 23.2\,\mathrm{dB}$ and instantaneous bandwidth $B = 9.2\,\mathrm{MHz}$, which we extract from individual Lorentzian fits to both lobes [cf. Fig.\,\ref{fig_noise_visibility}a) red line]. By monitoring the output power spectrum with and without the pump tone applied, we observe a maximal noise visibility, i.e. noise rise compared to the background outside the band, of $\Delta P = 14.2\,\mathrm{dB}$ at the frequency of maximum power gain [cf. Fig.\,\ref{fig_noise_visibility}a) blue line]. This implies that the HEMT noise only accounts for $4\,\%$ of the room temperature noise. In Fig.\,\ref{fig_noise_visibility}b) we show the measured gain-bandwidth (GB) product $\sqrt{G_0} B$ [cf. left hand panel] and noise visibility $\Delta P$ [cf. right hand panel] at maximum power gain $G_0$ for various pump strength. The GB-product is in good agreement with the theoretical prediction indicated by the black dashed line \cite{Eichler14_DPA}.  
\subsection{Measurement efficiency}
\begin{figure}[!t]
\begin{center}
\includegraphics[width = 1\columnwidth]{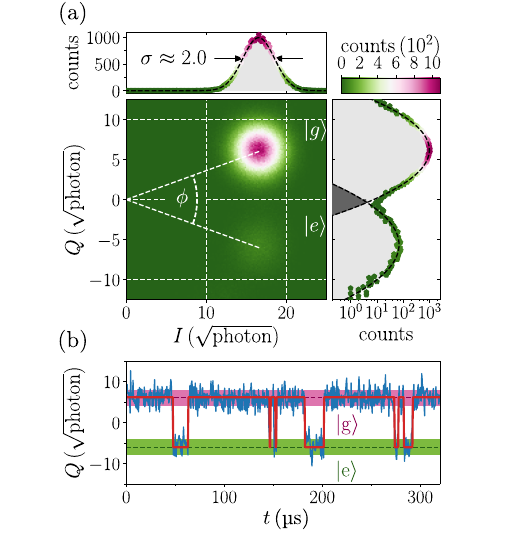}
\caption{\textbf{Quantum jumps measurement.} \textbf{a)} 2D-histrogram of measured $I$ and $Q$ quadratures of the readout resonator at $f_\mathrm{r} = 5.8224\,\mathrm{GHz}$. The $I$ and $Q$ values are reported in units of square root of measurement photons $\sqrt{\bar{n}_\mathrm{r} \kappa_\mathrm{r} T_\mathrm{m} / 4}$, where $\bar{n}_\mathrm{r} \approx 150$, $\kappa_\mathrm{r} / 2 \pi = 2.7\,\mathrm{MHz}$ and $T_\mathrm{m} = 500\,\mathrm{ns}$, for a total of $6 \times 10^5$ counts. We observe two peaks in the $IQ-$plane, a larger one corresponding to the qubit ground state $\ket{g}$, and an approximately ten times smaller peak corresponding to the first excited state $\ket{e}$. The qubit temperature calculated from the observed populations is $T_\mathrm{q} = 87\,\mathrm{mK}$, in agreement with the base plate temperature of the dilution refrigerator of $80\,\mathrm{mK}$, which was raised to activate thermal excitations of the qubit. The distribution plots along $I$ (top panel) and $Q$ (right panel) are plotted for slices centered on the ground state peak. From a Gaussian-fit we extract a standard deviation $\sigma = 2.0\,\mathrm{\sqrt{photon}}$.  \textbf{b)} Typical example of a measured quantum jump trace corresponding to the time evolution of the $Q$ quadrature for the same experiment shown in a). The red solid line indicates the qubit state given by a two point latching filter. The colored areas around the $Q$ values corresponding to $\ket{g}$ (pink) and $\ket{e}$ (green) represent one standard deviation $\sigma$, which is the value used for the latching filter.}
\label{fig_Quantum_jumps}
\end{center}
\end{figure}

To characterize the measurement efficiency $\eta$ of the setup, we measure the noise added by the DJJAA by decomposing the output signal into its quadratures $I(t)$ and $Q(t)$. In order to calibrate the measurement photon number $n_\mathrm{meas}$ [cf. Eq.\,\ref{EQ:nmeas}] corresponding to the recorded quadrature voltages, we calibrate the mean circulating number of photons inside the resonator $\bar{n}_\mathrm{r}$ by measuring the qubit's fundamental transition frequency $f_\mathrm{q}$ in a sequence of Ramsey fringes experiments \cite{Ramsey50}. The frequency $f_\mathrm{R}$ of the Ramsey oscillations is given by the drive detuning from the qubit's transition frequency $f_\mathrm{R} = |f_\mathrm{d} - f_\mathrm{q}|$, which is chosen to be comparable to the frequency shift induced by the population of the readout resonator $\Delta f_\mathrm{q} = \bar{n}_\mathrm{r} \chi_\mathrm{qr} $, where $\chi_\mathrm{qr} = 480\,\mathrm{kHz}$ is the qubit dispersive shift \cite{Blais04}. We populate the resonator with $\bar{n}_\mathrm{r}$ photons by simultaneously applying a constant tone to the readout resonator at frequency $f_\mathrm{r}$ [cf. App.\,\ref{A_photon_number}].\\
\\
As shown in Fig.\,\ref{fig_Quantum_jumps}, we can now express the histogram of measured signal quadratures $I(t)$ and $Q(t)$ in units of the measurement photon amplitude $\sqrt{n_\mathrm{meas}}$, for $\bar{n}_\mathrm{r} \approx 150$ and an integration time $T_\mathrm{m} = 500\,\mathrm{ns}$. The qubit's ground state $\ket{g}$ and first excited state $\ket{e}$ are visible as two circles with Gaussian profile. For clarity, we rotate the $IQ$-plane such that the information about the qubit state is encoded entirely in the $Q$-quadrature, as shown by the slices through the histogram along $I$ [cf. top panel Fig.\,\ref{fig_Quantum_jumps}a)] and $Q$ [cf. right panel Fig.\,\ref{fig_Quantum_jumps}a)]. The angle between the ground and first excited state is $\phi = 4 \arctan \left( \chi_\mathrm{qr} / \kappa \right) \approx 40^\circ$. \\
\\
We calculate the measurement efficiency $\eta$ at the resonance frequency of the readout resonator by comparing the measured standard deviation $\sigma = 2.0\,\sqrt{\mathrm{photon}}$ of the ground state distribution in our histogram, with the ideal case, which is $\sigma_\mathrm{ideal} = 1 / \sqrt{2}\,\sqrt{\mathrm{photon}}$ for a coherent state \cite{Caves1982}, and we obtain $\eta = \sigma_\mathrm{ideal}^2 / \sigma^2 \approx 0.13$. \\
\\
We can partly attribute the reduction in measurement efficiency to known and expected losses between the readout resonator and the parametric amplifier. Since we use a commercial power combiner to feed in the pump tone, we lower the power incident on the DJJAA by at least a factor of two compared to the calibrated value. Additionally, our sample holder is separated from the amplifier by several microwave components, such as circulators and various connecting cables, which can further add dissipation at the readout frequency. Therefore, a conservative upper bound on the measurement efficiency determined by losses is $\eta_\mathrm{L} \leq 0.5$, which in combination with the measurement efficiency due to the HEMT, implies a conservative bound on the quantum efficiency $\eta_\mathrm{DJJAA} = \eta /  \eta_\mathrm{L} \geq 0.26$ for our parametric amplifier.\\
\\
An example of a measured quantum jump trace is depicted in Fig.\,\ref{fig_Quantum_jumps}b) for the same integration time and readout strength as in a). For the detection of the qubit state, highlighted by the solid red line, we use a multi-point filter \cite{Vool14}. The filter declares a jump, when the detected $Q$-value falls into a range of $\pm \sigma$ around the mean values $\bar{Q}_\mathrm{i}$ ($i \in \mathrm{g,e,f,h}$) associated with the first four qubit states. For the ground and first excited state, the filter-range is color coded by the pink and green areas around the mean values, respectively, which are indicated by dashed lines.  As discussed in App.\,\ref{A_qubit_temperature}, from the relative population of the first four qubit states, we extract a qubit temperature of $T_\mathrm{q} = 87\,\mathrm{mK}$, which is in good agreement to the temperature of the cryostat base plate $T \approx 80\,\mathrm{mK}$ during this experiment. For measurements taken at base temperature ($T_\mathrm{base} = 30\,\mathrm{mK}$), the qubit temperature saturates at a constant value $T_\mathrm{q} \approx 61 \, \mathrm{mK}$.

\section{Conclusion}
\label{SEC:conclusion}
In summary, we have demonstrated a type of parametric amplifier based on a dispersion engineered Josephson junction array, with up to four hybridized pairs of modes, showing non-degenerate power gain in excess of $20\,\mathrm{dB}$, within an instantaneous bandwidth in the range of $10\,\mathrm{MHz}$. The measured noise visibility suggests that our amplifiers approach the quantum limit of added noise. We used the amplifier to measure quantum jumps of a transmon qubit, which we can discriminate with $90\%$ fidelity within $500\,\mathrm{ns}$ of integration. For comparison, using state-of-the-art readout chain optimization and pulse shaping, fidelities up top $97\,\%$ can be achieved in $80\,\mathrm{ns}$ \cite{Heinsoo18}. The optical fabrication of the JJ array is accessible, reproducible, and low cost. 

By optimizing the mode engineering of future amplifiers it will be possible to build an array with sufficiently high mode density, such that a moderate flux tunability will suffice to cover the entire frequency band from $1\,\mathrm{GHz}$ up to $10\,\mathrm{GHz}$. Further improvements also include the suppression of unintended higher-order effects by utilizing asymmetric dc-SQUIDs \cite{Eichler14_DPA} or SNAILs \cite{SivJAMPA19, Frattini17, Siv19}. \\

Funding was provided by the Alexander von Humboldt foundation in the framework of a Sofja Kovalevskaja award endowed by the German Federal Ministry of Education and Research,  and by the Initiative and Networking Fund of the Helmholtz Association, within the Helmholtz Future Project \textit{Scalable solid state quantum computing.} Furthermore, this research was supported by the ANR under contracts CLOUD (project number ANR-16-CE24-0005). PW and WW acknowledge support from the European Research Council advanced grant MoQuOS (N. 741276). IT and AVU acknowledge partial support from the Ministry of Education and Science of the Russian Federation in the framework of the Increase Competitiveness Program of the National University of Science and Technology MISIS (Contract No. K2-2018-015). Facilities use was supported by the  KIT  Nanostructure  Service  Laboratory  (NSL).  We acknowledge qKit \cite{qkit} for providing a convenient measurement software framework.

\bibliography{DJJAA_references}



\appendix

\renewcommand{\appendixname}{Appendix}

\section{Inductance and capacitance matrix}
\label{A_inductance_matrix}
The capacitance and inverse inductance matrices introduced in Eq.\,\ref{LAG:flux}, $\tilde{C}$ and $\tilde{L}^{-1}$, respectively, are
\onecolumngrid
\begin{equation}
\tilde{C} =
  \begin{pmatrix}
   			2 C_\mathrm{J} + C_0 & -C_\mathrm{J} & 0 & \dots & & & & & &  \\
-C_\mathrm{J} & 2 C_\mathrm{J} + C_0 & -C_\mathrm{J} & 0 & \dots & & & & & \\
0 & -C_\mathrm{J} & 2 C_\mathrm{J} + C_0 & -C_\mathrm{J} & 0 & \dots & & & & \\
\vdots & \ddots & \ddots & \ddots & \ddots & \ddots & \ddots & & & \\
 & & 0 & -C_\mathrm{J} & \textcolor{red}{C_\mathrm{J} + C_\mathrm{c} + C_0^\prime} & \textcolor{red}{-C_\mathrm{c}} & 0 & & &  \\
 & & & 0 & \textcolor{red}{-C_\mathrm{c}} & \textcolor{red}{C_\mathrm{J} + C_\mathrm{c} + C_0^\prime} & -C_\mathrm{J} & 0 & & \\
 & & & & 0 & -C_\mathrm{J} & 2 C_\mathrm{J}  + C_0 & -C_\mathrm{J} & 0 &  \\
  &   &  &  &   \ddots & \ddots & \ddots & \ddots & \ddots & \vdots\\
  &   &  &  &  & \ddots & 0 & - C_\mathrm{J} & 2 C_\mathrm{J} + C_0 & -C_\mathrm{J}\\
  &  &  &  &  &  & \dots & 0 & - C_\mathrm{J} & 2 C_\mathrm{J} + C_0 \\
  \end{pmatrix} \\
\end{equation}
and
\begin{equation}
\tilde{L}^{-1} =
\begin{pmatrix}
   			\frac{2}{L_\mathrm{J}} & -\frac{1}{L_\mathrm{J}} & 0 & \dots & & & & & &  \\
-\frac{1}{L_\mathrm{J}} & \frac{2}{L_\mathrm{J}} & -\frac{1}{L_\mathrm{J}} & 0 & \dots & & & & & \\
0 & -\frac{1}{L_\mathrm{J}} & \frac{2}{L_\mathrm{J}} & -\frac{1}{L_\mathrm{J}} & 0 & \dots & & & & \\
\vdots & \ddots & \ddots & \ddots & \ddots & \ddots & \ddots & & & \\
 & & 0 & -\frac{1}{L_\mathrm{J}} & \textcolor{red}{\frac{1}{L_\mathrm{J}}} & \textcolor{red}{0} & 0 & & &  \\
 & & & 0 & \textcolor{red}{0} & \textcolor{red}{\frac{1}{L_\mathrm{J}}} & -\frac{1}{L_\mathrm{J}} & 0 & & \\
 & & & & 0 & -\frac{1}{L_\mathrm{J}} & \frac{2}{L_\mathrm{J}} & -\frac{1}{L_\mathrm{J}} & 0 &  \\
  &   &  &  &   \ddots & \ddots & \ddots & \ddots & \ddots & \vdots\\
  &   &  &  &  & \ddots & 0 & -\frac{1}{L_\mathrm{J}} & \frac{2}{L_\mathrm{J}} & -\frac{1}{L_\mathrm{J}}\\
  &  &  &  &  &  & \dots & 0 & -\frac{1}{L_\mathrm{J}} & \frac{2}{L_\mathrm{J}} \\
\end{pmatrix}.
\end{equation}
\twocolumngrid
The parameters for the SQUID array are the single SQUID Josephson inductance $L_\mathrm{J}$ and Josephson capacitance $C_\mathrm{J}$ and the capacitance per island to ground $C_0$. The matrix elements due to the center capacitance $C_\mathrm{c}$ and its capacitance to ground $C_\mathrm{0}^\prime$ are highlighted in red. Since the array is galvanically connected to its environment on both ends, the boundary conditions are $\Phi_{0} = \Phi_{N+1} = 0$. As discussed in the main text, we are not considering long-range Coulomb interactions in the capacitance matrix that could be mediated by the ground plane on the backside of the sapphire wafer. 

\begin{table*}[t]
\begin{center}
\begin{tabular}{||c || c c c || c c c || c c c||} 
\hline
$N$ & \multicolumn{3}{c||}{1200} & \multicolumn{3}{c||}{1600} & \multicolumn{3}{c||}{1800} \\ \hline \hline
$m$ & $\dfrac{\omega_{m}}{2 \pi}$ & $\dfrac{K_{m,m}}{2 \pi}$ & $\dfrac{K_{m,m+1}}{2 \pi}$ & $\dfrac{\omega_{m}}{2 \pi}$ & $\dfrac{K_{m,m}}{2 \pi}$ & $\dfrac{K_{m,m+1}}{2 \pi}$ & $\dfrac{\omega_{m}}{2 \pi}$ & $\dfrac{K_{m,m}}{2 \pi}$ & $\dfrac{K_{m,m+1}}{2 \pi}$ \\
 & (GHz) & (kHz) & (kHz) & (GHz) & (kHz) & (kHz)& (GHz) & (kHz) & (kHz) \\  \hline
0 & 2.061 & 1.1 & 2.8 & 1.113 & 0.5 & 1.2 & 0.863 & 0.3 & 0.9 \\ 
1 & 2.478 & 1.8 & 6.1 & 1.345 & 0.8 & 2.7 & 1.039 & 0.6 & 1.9 \\ 
2 & 6.427 & 12.7 & 26.9 & 3.488 & 5.6 & 11.9 & 2.696 & 3.8 & 8.2 \\ 
3 & 7.106 & 15.3 & 29.7 & 3.927 & 7.1 & 13.7 & 3.050 & 5.0 & 9.5 \\ 
4 & 10.398 & 34.9 & 69.8 & 5.828 & 16.4 & 32.8 & 4.538 & 11.5 & 22.8 \\ 
5 & 10.881 & 37.3 & 59.6 & 6.210 & 18.2 & 29.4 & 4.873 & 12.9 & 20.9 \\ 
6 & 13.364 & 58.3 & 115.3 & 7.819 & 30.0 & 58.7 & 6.177 & 21.5 & 41.9 \\ 
7 & 13.646 & 59.8 & 87.2 & 8.090 & 31.4 & 46.5 & 6.434 & 22.8 & 33.9 \\ 
8 & & & & 9.381 & 43.3 & 84.7 & 7.528 & 32.1 & 62.4 \\ 
9 & & & & 9.560 & 44.3 & 62.4 & 7.710 & 33.1 & 46.8 \\ 
10 & & & & 10.561 & 55.0 & 107.8 & 8.598 & 42.0 & 81.8 \\ 
11 & & & & 10.677 & 55.7 & 76.1 & 8.724 & 42.7 & 58.5 \\ \hline
\end{tabular}
\end{center}
\caption{Eigenfrequencies, self- and neighbouring modes cross-Kerr coefficients $\omega_m$, $K_{m,m}$ and $K_{m, m+1}$, respectively, calculated with our circuit model [cf. Sec\,\ref{SEC:CircuitModel}] for the the first twelve eigenmodes of the three devices discussed in Sec.\,\ref{SEC:Gainmeasurements}. The parameter values for the circuit model are found in Tab.\,\ref{tab:sampleoverview} of App.\,\ref{A_overview} for all three devices.}
\label{tab:KERR}
\end{table*}

\section{Non-linearity: self- and cross-Kerr coefficients}
\label{A_Kerr}   
For Josephson parametric amplifiers (JPA), the non-linearity arising from the cosine potential of the JJ(s) is an indispensable requirement for the parametric amplification process. However, as discussed in \cite{Boutin17,Liu17}, the magnitude of the lowest order non-linear terms, referred to as Kerr coefficients, influences the amplifier's saturation power and the required parametric pump power at the same time. Therefore it is important to derive the non-linearity for the various eigenmodes of our circuit design as a function of the circuit parameters. As described in \cite{Weissl15}, we introduce the non-linearity pertubatively to our linear circuit model. The modal distribution of the eigenmodes enters the two equations for the self- and cross-Kerr coefficients 
\begin{equation}
\begin{split}
K_{m,m} &=  \frac{2 \hbar \pi^4 E_\mathrm{J} \eta_{mmmm}}{\Phi_0^4 C_\mathrm{J}^2 \omega_\mathrm{m}^2}\\
K_{m,k} &= \frac{4 \hbar \pi^4 E_\mathrm{J} \eta_{mmkk}}{\Phi_0^4 C_\mathrm{J}^2 \omega_\mathrm{m} \omega_\mathrm{k}}
\end{split}
\label{EQ:KERR_app}
\end{equation}
by a dimensionless factor $\eta_{mmkk}$.  
\begin{equation}
\begin{split}
\eta_{mmkk} = & \sum_{i = 0}^{N} \left[ \left( \sum_{j = 0}^{N} \left( \sqrt{C_\mathrm{J}} \tilde{C}^{-1/2}_{i,j} - \sqrt{C_\mathrm{J}} \tilde{C}^{-1/2}_{i-1,j}\right) \Psi_{j,m} \right)^2 \right.  \\
& \left. \times \left( \sum_{j = 0}^{N} \left( \sqrt{C_\mathrm{J}} \tilde{C}^{-1/2}_{i,j} - \sqrt{C_\mathrm{J}} \tilde{C}^{-1/2}_{i-1,j}\right) \Psi_{j,k} \right)^2 \right]
\end{split}
\end{equation}
Here $\tilde{C}^{-1/2}$ is the square root of the inverse capacitance matrix, defined as $\tilde{C}^{-1/2} \cdot \tilde{C}^{-1/2} = \tilde{C}^{-1}$, and $\Psi_{j,m}$ is the $j$-th entry of the $m$-th eigenvector $\vec{\Psi}_m$ of the eigenvalue problem Eq.\,\ref{EQ:EIGENPROBLEM}. \\
Generally speaking, the more two eigenmodes overlap in terms of their standing-wave flux distribution along the array, the stronger they influence each other when driven. Notably, for the antisymmetric modes (even mode number $n$), the jump in the node phase and node flux distribution, which occurs from negative to positive values at the position of the center capacitance, causes unphysical results for the dimensionless factors $\eta_{mmmm}$ and, as a consequence, artificially large self-Kerr coefficients. Therefore, we calculate the self- and neighbouring modes cross-Kerr coefficients for our devices by symmetrizing these eigenfunctions, which does not alter the physical meaning but avoids the numercial error. 
\begin{equation}
\vec{\Psi}_{m, \mathrm{sym}} = \vec{\Psi}_{m}^\mathrm{T} \cdot \tilde{S}
\end{equation}
Here, the diagonal matrix $\tilde{S}$ is 
\begin{equation}
\tilde{S} = \mathrm{diag}(1,1,...,1,1,-1,-1,...,-1,-1),
\end{equation}
where the change in sign occurs at the entry with index $j = N/2 + 1$.

\section{Circuit model: device overview}
\label{A_overview}
In the main text we discuss three samples, which slightly differ in their design parameters. In Tab.\,\ref{tab:sampleoverview}, we give an overview of the circuit parameters that enter our model. \\
We evaluate the overlap areas $A_\mathrm{J}$ forming the JJ contact using SEM images. We didn't investigate the devices directly, in order to avoid damaging the sample. Instead we used comparable samples from the same batch in the immediate vicinity of the sample chip location. The Josephson capacitance is inferred from the overlap area: $C_\mathrm{J} = 50\,\mathrm{fF/\si{\micro\metre^2}} \times A_\mathrm{J}$. The Josephson inductance per SQUID junction $L_\mathrm{J}$ is deduced from room temperature resistance measurements. The value of the center capacitance is estimated from finite-element simulations. The capacitance to ground of the center capacitor plates are calculated for a standard microstrip geometry with backside metallization \cite{Paul94}. Therefore, the only remaining free circuit parameter is the capacitance per ground $C_0$ of each superconducting island in between two SQUID junctions, which we obtain from fitting the dispersion relation [cf. Fig.\,\ref{fig_circuit_model}]. The stray inductance $L_\mathrm{s}$ of the aluminum leads is estimated from the fits presented in App.\,\ref{A_FluxMod}. Its contribution to the coupling rate $\kappa$ to the input port is discussed in App.\,\ref{A_Envi}, and, for simplicity, it is neglected in the linear model used to calculate the dispersion relation [cf. Fig.\,\ref{fig_circuit_model}]. 
\begin{table}[h!]
\begin{center}
\begin{tabular}{||c | c  | c | c||} 
\hline
& sample I & sample II & sample III \\ \hline \hline
N & 1200 & 1600 & 1800 \\ 
$R_{\mathrm{n, in}}\,\mathrm{(k\Omega)} $ & 29.03 & 70.96 & 102.25 \\
$R_{\mathrm{n, out}}\,\mathrm{(k\Omega)} $ & 28.41 & 71.3 & 102.91 \\
$m$ & 1.022 & 0.995 & 0.994 \\
$R_\mathrm{n,SQ}\,\mathrm{(\Omega)}$ & 41.5 & 82.5 & 108 \\
$I_\mathrm{c}\,\si{(\micro\ampere)}$ & 6.0 & 3.0 & 2.3 \\
$L_\mathrm{J}\,\mathrm{(pH)}$ & 55 & 110 & 143 \\ 
$L_\mathrm{s}\,\mathrm{(pH)}$ & 12.6 & 12.6 & 13.3 \\
$\gamma_\mathrm{L, fit}$ & 0.67 & 0.87 & $0.92 \pm 2$ \\
$\gamma_\mathrm{L}$ & 0.81 & 0.90 & 0.91 \\
$A_\mathrm{J}\,\si{(\micro\metre^2)}$ & $10.8 \pm 0.4$ & $10.5 \pm 0.4$ & $10.5\pm 0.4$ \\
$C_\mathrm{J}\,\mathrm{(fF)}$ & $1080\pm 40$ & $1050 \pm 40$ & $1050 \pm 40$ \\
$C_\mathrm{c}\,\mathrm{(fF)} $ & 30 & 40 & 45 \\
$C_\mathrm{0}^\prime\,\mathrm{(fF)}$ & 33 & 33 & 33 \\
$C_0\,\mathrm{(fF)}$ & 0.39 & 0.40 & 0.42 \\ 
$\omega_\mathrm{pl}/2 \pi\,\mathrm{(GHz)}$ & 20.65 & 14.81 & 13.0 \\\hline
\end{tabular}
\end{center}
\caption{Overview of the device circuit parameters: number of SQUID junctions $N$, room temperature resistance of the array from the input port to the center capacitance $R_\mathrm{n, in}$, room temperature resistance of the array from the center capacitance to ground $R_\mathrm{n, out}$, resistance asymmetry between the two array sections  $m$, room temperature resistance of a single SQUID junction $R_\mathrm{n,SQ}$, critical current of a single SQUID junction $I_\mathrm{c,SQ}$, Josephson inductance of a single SQUID junction $L_\mathrm{J}$,  stray inductance $L_\mathrm{s}$, kinetic inductance participation ratio $\gamma_\mathrm{L, fit}$ [cf.\,App.\ref{A_FluxMod}] and $\gamma_\mathrm{L} = L_\mathrm{J} / (L_\mathrm{J} + L_\mathrm{s})$, Junction area $A_\mathrm{J}$, Josephson capacitance $C_\mathrm{J}$, center capacitance $C_\mathrm{c}$, capacitance to ground $C_\mathrm{0}^\prime$ (center capacitance), capacitance to ground per island $C_0$, plasma frequency $\omega_\mathrm{pl}$.} 
\label{tab:sampleoverview}
\end{table}

\begin{figure*}[t]
\begin{center}
\includegraphics[width=6.69in]{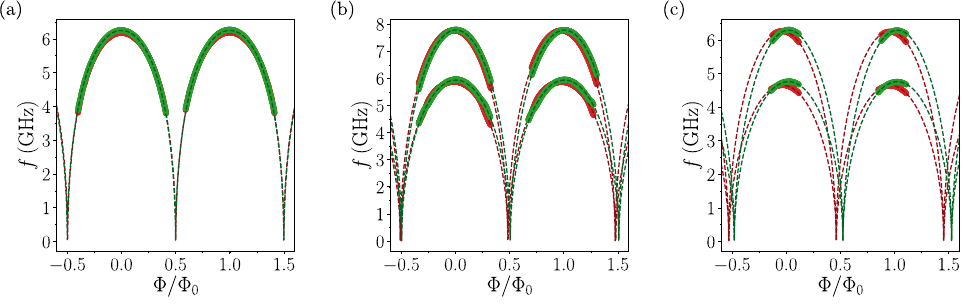}
\caption{Fitted uncoupled eigenfrequencies (i.e. $J = 0$) of the three devices discussed in the main text, \textbf{a)} $N = 1200$, dimer number $n = 2$, \textbf{b)} $N = 1600$, $n \in  \{3, 4\}$, \textbf{c)} $N = 1800$, $n \in  \{3, 4\}$, as a function of the effective external flux $\Phi$. The data points are extracted from the measurement data depicted in Fig.\,\ref{fig_gain_measurements} by fitting Eq.\,\ref{EQ:GAMMA_REF} to the frequency dependence of the reflection coefficient and following the calculation described in App.\,\ref{A_Dimer}. The color coded dashed lines are the fit results according to Eq.\,\ref{EQ:omega_fit}, from which we extract the Josephson inductance participation ratio $\gamma_\mathrm{L}$. In abscence of field offsets and asymmetries in the array, the red and green curves should perfectly overlap. While the first device operates in a rather uniform magnetic environment, the latter two see an external magnetic stray field gradient. However, only for the last sample the stray field gradient has an influence on the hybridization, since the coupling between neighbouring modes $J_\mathrm{n}$ is weaker.}
\label{supfig_participation_ratio}
\end{center}
\end{figure*}

\section{Flux modulation and Josephson inductance participation ratio}
\label{A_FluxMod}
The modulation of the system eigenmodes $\omega_\mathrm{n}$ as a function of the bias current $I_\mathrm{b}$ applied to the superconducting field coil, can be described by an effective lumped-element model for each mode, which consists of a series circuit of a flux-dependent and flux-independent inductance $L_\mathrm{J,tot} (\Phi)$ and $L_\mathrm{S}$, respectively, shunted by a capacitance $C$. The effective resonance frequency of the lumped-element model writes
\begin{equation}
f \left( I_\mathrm{b} \right) = \frac{1}{2 \pi \sqrt{C \left[L_\mathrm{S} + L_\mathrm{J,tot} (I_\mathrm{b}) \right]}}.
\label{EQ:omega_eff}
\end{equation}
While $L_\mathrm{J,tot}$ originates from the kinetic inductance of the SQUIDs, the effective stray inductance $L_\mathrm{s}$ originates from both, geometric and kinetic contributions of the superconducting leads connecting the SQUIDs. By fitting the model prediction to the experimentally observed frequency modulation, we can calculate the Josephson inductance participation ratio $\gamma_\mathrm{L} = L_\mathrm{J,tot} / \left(L_\mathrm{J,tot} + L_\mathrm{S}\right)$, and also calibrate the effective bias flux $\Phi = \Phi_\mathrm{b} + \Phi_\mathrm{offset}$ in units of $\Phi_0$. The bias flux $\Phi_\mathrm{b}$ is due to the current $I_\mathrm{b}$ applied to the coil, and $\Phi_\mathrm{offset}$ is a static offset. \\
\\
The current dependence of $L_\mathrm{J,tot}$ is deduced from the flux dependent inductance of a single symmetric DC-SQUID, which is given by
\begin{equation}
L_\mathrm{J,tot} \left( \Phi \right) = \frac{L_\mathrm{J,tot}(0)}{  \left|\cos \left( \frac{\pi \Phi}{\Phi_0}\right)\right| },
\label{EQ:LKSQUID_sym}
\end{equation} 
Assuming a linear dependence between the external bias flux created by the superconducting coil and the bias current, Eq.\,\ref{EQ:LKSQUID_sym} is expressed as a function of bias current
\begin{equation}
L_\mathrm{J,tot} \left( I_\mathrm{b} \right) = \frac{L_\mathrm{J,tot} (- I_\mathrm{offset})}{ \left|\cos \left( \pi l_\mathrm{b} (I_\mathrm{b} + I_\mathrm{offset}\right)\right|}.
\label{EQ:LKCurrent}
\end{equation}
The parameter $l_\mathrm{b}$ translates the applied current bias into an external flux bias expressed in number of magnetic flux quanta. The parameter $I_\mathrm{offset}$ accounts for the presence of an offset flux due to magnetic stray fields produced by the attached circulator and other components of the setup, or the earth's magnetic field. \\
Inserting Eq.\,\ref{EQ:LKCurrent} into Eq.\,\ref{EQ:omega_eff}, the fit-function for the current modulation of the resonance frequencies writes 
\begin{equation}
f \left( I_\mathrm{b} \right) = \frac{1}{2 \pi \sqrt{C \left[L_\mathrm{S} + L_\mathrm{J,tot}(- I_\mathrm{offset}) / \left|\cos \left( \pi l_\mathrm{b} (I_\mathrm{b} + I_\mathrm{offset}\right)\right|  \right]}}.
\label{EQ:omega_gamma}
\end{equation}
By inserting the Josephson inductance participation ratio $\gamma_\mathrm{L} = L_\mathrm{J,tot}(0) / \left(L_\mathrm{S} +  L_\mathrm{J,tot} (0) \right)$ into Eq.\,\ref{EQ:omega_gamma} and introducing the frequency amplitude $f_\mathrm{0} = 1 / 2 \pi \sqrt{C \left(L_\mathrm{S} + L_\mathrm{J,tot}(0) \right)}$, we obtain the final expression for the fitting function used in Fig.\,\ref{fig_gain_measurements} and Fig.\,\ref{supfig_participation_ratio}
\begin{equation}
f \left( I_\mathrm{b} \right) = \frac{f_\mathrm{0}}{\sqrt{1 + \gamma_\mathrm{L} \left[1 + \left|\cos \left( \pi l_\mathrm{b} (I_\mathrm{b} + I_\mathrm{offset}\right)\right|^{-1} \right]}}.
\label{EQ:omega_fit}
\end{equation}
\begin{figure*}[t!]
\begin{center}
\includegraphics[width=6.69in]{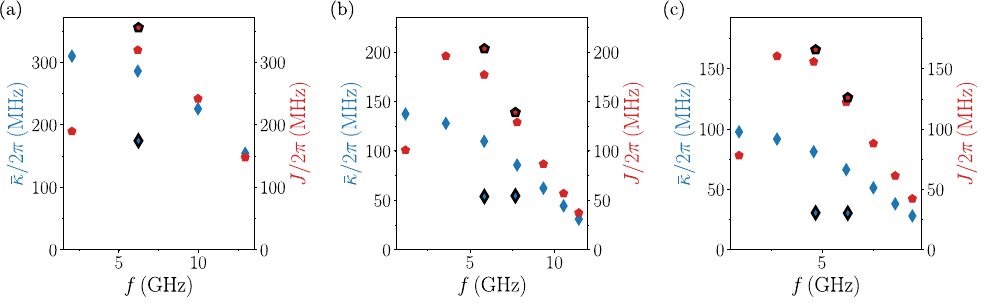}
\caption{Calculated external coupling rate $\kappa$ (blue diamonds) and dimer mode splitting $J$ (red pentagons) as a function of the mode frequency predicted by our transmission matrix model. The corresponding circuit parameters used for the calculation are listed in Tab.\,\ref{tab:sampleoverview}. The experimental data is depicted with an additional black frame and is extracted at the flux sweet spot of each mode by fitting Eq.\,\ref{EQ:GAMMA_REF} to the measured complex reflection coefficient in close vicinity. The results presented in panels \textbf{a) - c)} correspond to the samples discussed in the main text [cf. Fig.\,\ref{fig_gain_measurements}], $N = 1200$, $N = 1600$ and $N = 1800$, respectively. }
\label{supfig_coupling_environment}
\end{center}
\end{figure*}
For each array, we extract the Josephson inductance participation ratio $\gamma_\mathrm{L}$ between 0.67 and 0.92, as listed in Tab.\,\ref{tab:sampleoverview}. These values are consistent with estimates based on the geometry and the thickness of the aluminum film, for which we expect a stray kinetic sheet inductance of $1.3 \, \mathrm{pH / \square}$ and a stray geometric inductance \cite{Paul94} of $1.5 \, \mathrm{pH / \si{\micro\metre}}$, amounting of a total stray inductance per SQUID of $15 \pm 4\,\mathrm{pH}$.

\section{Coupling to the environment}
\label{A_Envi}
The coupling strength $\kappa$ of the amplifier modes to the input port, which is implemented with a transmission line of characteristic impedance $Z_0 = 50\,\mathrm{\Omega}$, is calculated with a transmission matrix approach \cite{Poz11}. In this formalism, the total transmission matrix $\tilde{T}$ of the amplifier is calculated by multiplying the individual transmission matrices $\tilde{T}_i$ associated with each circuit element, e. g. the SQUIDs, superconducting islands and stray inductances.
\begin{equation}
\tilde{T} = \prod_{i = 1}^{N + 1} \tilde{T}_i =
	\begin{pmatrix}
    A & B \\
    C & D
  \end{pmatrix}
\end{equation}
Here, $N$ is the total number of SQUIDs. Notably, the sum runs over $N + 1$ entries due to the additional site introduced by the center capacitance $C_\mathrm{c}$. The individual circuit elements expressed in transmission matrices are 
\begin{equation}
\begin{split}
\tilde{T}_\mathrm{LS} &=
  \begin{pmatrix}
    1 & j \omega L_\mathrm{stray}  \\
    0 & 1
  \end{pmatrix} \\
\tilde{T}_\mathrm{SQ} &=
  \begin{pmatrix}
    1 & \left[ \left(j \omega L_\mathrm{J} \right)^{-1} + j \omega C_\mathrm{J}  \right]^{-1} \\
    0 & 1
  \end{pmatrix} \\
\tilde{T}_\mathrm{C_0} &=
  \begin{pmatrix}
    1 & 0 \\
    {j \omega C_0} & 1
  \end{pmatrix} \\
\tilde{T}_\mathrm{C_\mathrm{c}} &=
  \begin{pmatrix}
    1 & 1 / {j \omega C_\mathrm{c}} \\
    0 & 1
  \end{pmatrix} \\
\tilde{T}_\mathrm{C_0^\prime} &=
  \begin{pmatrix}
    1 & 0 \\
     {j \omega C_0^\prime} & 1
  \end{pmatrix} \\
 \end{split}.
\end{equation}
Note that $j$ is the imaginary unit according to electrical engineering standards ($j = - i$), $L_\mathrm{stray}$ is the stray inductance due to the pure aluminum islands, $L_\mathrm{J}$ and $C_\mathrm{J}$ are the Josephson inductance and capacitance for each SQUID junction, respectively, $C_0$ is the island capacitance to ground, $C_\mathrm{c}$ is the center capacitance and $C_\mathrm{0}^\prime$ is the capacitor plates' capacitance to ground. The total transmission matrix of our amplifiers is composed of a repetition of $N$ times three elements ($\tilde{T}_\mathrm{LS}$, $\tilde{T}_\mathrm{SQ}$ and $\tilde{T}_\mathrm{c_0}$), with the contribution of the center capacitance ($\tilde{T}_\mathrm{C_0^\prime}$ and $\tilde{T}_\mathrm{C_c}$) right in the middle 
\begin{equation}
\tilde{T} = \left( \prod_{i = 1}^{N/2} \tilde{T}_\mathrm{LS}  \tilde{T}_\mathrm{SQ}  \tilde{T}_\mathrm{c_0} \right) \tilde{T}_\mathrm{C_0^\prime} \tilde{T}_\mathrm{C_\mathrm{c}} \tilde{T}_\mathrm{C_0^\prime} \left( \prod_{i = 1}^{N/2} \tilde{T}_\mathrm{LS}  \tilde{T}_\mathrm{SQ}  \tilde{T}_\mathrm{c_0} \right).
\end{equation}
From the total transmission matrix $\tilde{T}$, the complex reflection coefficient $S_\mathrm{11}$ is calculated
\begin{equation}
S_\mathrm{11} = \frac{A + B / Z_\mathrm{0} - C Z_0 - D}{A + B / Z_0 + C Z_0 + D}.
\label{EQ:Scattering}
\end{equation} 
For the input port, we assume a characteristic impedance $Z_0 = 50\,\mathrm{\Omega}$. By solving Eq.\,\ref{EQ:Scattering} in the frequency range up to the plasma frequency $\omega \in \left[0 , \omega_\mathrm{pl} \right ]$ numerically, we can extract the eigenfrequencies, external coupling rate $\kappa_\mathrm{n}$ and level splitting $J_\mathrm{n}$ for each dimer individually. Figure\,\ref{supfig_coupling_environment} depicts the numerical calculation and the extracted experimental data (black contour markers) for the three samples discussed in the main text. As also stated in the main text, the coupling rate $J_n$ shows a dome like structure, while the coupling rate continuously decreases.

\begin{figure*}[t]
\begin{center}
\includegraphics[width=6.69in]{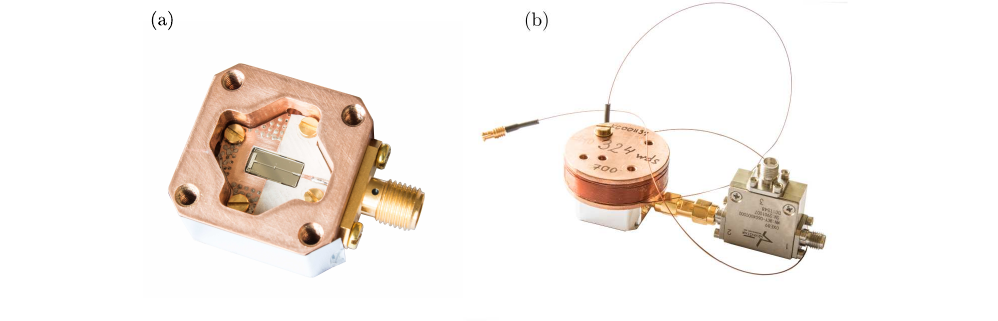}
\caption{\textbf{a)} Body of the copper sample holder (without lid), hosting a printed circuit board (PCB), with a $7.5 \times 3.6\,\mathrm{mm^2}$ sapphire wafer in the center. The PCB is glued with silver paste and attached with four brass screws. \textbf{b)} Closed copper sample holder, with a superconducting bias coil integrated into its lid, and a commercial circulator directly mounted to the SMA connector. }
\label{supfig_sample_holder}
\end{center}
\end{figure*}

\section{Sample holder}
\label{A_sample_holder}
The amplifiers, with a physical dimension of $7.5 \times 3.6\,\mathrm{mm^2}$, are glued into a copper sample holder with silver paste. For the external connection to the on-chip input port we use a printed circuit board (PCB) with a $50\,\mathrm{\Omega}$ microstrip transmission line [cf. Fig.\,\ref{supfig_sample_holder}a)]. The PCB is covered with copper on both sides, enclosing a low loss dielectric ($\epsilon_\mathrm{r} = 9.9$) of thickness $t = 635\,\si{\micro\metre}$. The height difference between the PCB and the sapphire wafer ($t = 330\,\si{\micro\metre}$) is compensated by a small copper post (not visible) below the sample. The PCB transmission line is directly soldered to the center conductor of an SMA connector. The other half of the PCB's top plate remains covered with copper and serves as ground for our amplifiers. For that reason, we use vias to galvanically connect the top copper plate to the sample holder. The amplifier is connected to the PCB at both ends with aluminum wire bonds. \\
Each amplifier is equipped with a superconducting bias coil integrated into the lid of the sample holder, and a commercial circulator directly mounted to the SMA input port, to avoid low-frequency standing waves [cf. Fig.\,\ref{supfig_sample_holder}b)].
\begin{figure*}
\begin{center}
\includegraphics[width=6.69in]{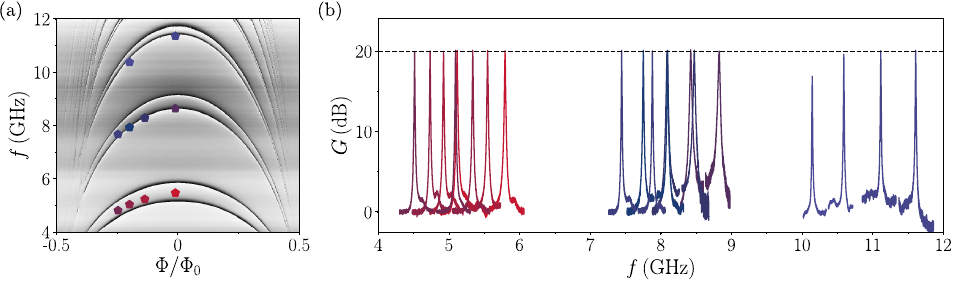}
\caption{\textbf{a)} Phase of the complex reflection coefficient $\arg (S_{11})$ in radians as a function of probe frequency $f$ and external bias flux $\Phi$, normalized to the flux quantum $\Phi_0$, for a DJJAA with $N = 1300$ SQUIDs. The greyscale covers the range from $-\pi$ (black) to $\pi$ (white). The main features in the given frequency range are the second ($n = 2$), third ($n =3$) and fourth ($n = 4$) dimer of the device, with higher frequency eigenmodes becoming visible close to full SQUID frustration ($\Phi / \Phi_0 = 0.5$). The  pentagons highlight the bias flux and pump frequency used to measure the power gain close to the dimer modes. \textbf{b)} Power gain $G$ in dB as a function of probe frequency $f$ for the bias flux and pump frequencies color-coded according to the pentagons in a). The tunable bandwidth of this device measured from the lowest frequency to the highest frequency of each dimer at which we reach $G = 20\,\mathrm{dB}$ exceeds $1.5\,\mathrm{GHz}$ for all three dimers. This is a typical value for our devices.}
\label{supfig_flux_tunability}
\end{center}
\end{figure*}
\section{Linear dimer characterization}
\label{A_Dimer}
In our devices, the frequency detuning between neighbouring dimers $\Delta \omega$ is much larger than the frequency detuning $2 J_n$ between the pair of modes that form a dimer. Therefore, we can neglect the other modes of the array when measruing a device in close vicinity of a dimer. In this limit, and assuming a weak probe tone, the measured reflection coefficient $\Gamma$ is simply the product of two modes with resonance frequency $\omega_+$ and $\omega_-$, external coupling rate $\kappa_+$ and $\kappa-$ to the input port, and internal loss rates $\gamma_+$ and $\gamma_-$. In the ideal case, without any variations along the arrays in terms of the critical current and external offset flux, the two external coupling rates are identical ($\kappa_+ = \kappa_-$). The dimer reflection coefficient is
\begin{equation}
\Gamma = \prod_{m \in \{+,- \}} \Gamma_m e^{i \phi_0},
\label{EQ:GAMMA_REF}
\end{equation}
where $\phi_0$ is an arbitrary offset-phase determined by the measurement setup, and $\Gamma_m$ is the standard reflection coefficient of a linear single-port resonator
\begin{equation}
\Gamma_m = - 1 + \frac{\frac{\kappa_m (\kappa_m + \gamma_m)}{2} + i \kappa_m (\omega - \omega_i)}{(\omega - \omega_0)^2 + \frac{(\kappa_m + \gamma_m)^2}{4}}.
\end{equation}
In the case that the external coupling rates $\kappa_\pm$ are not identical, we can gain information about the variation mainly in the Josephson energy along the array by perfoming a basis transformation. Instead of describing the system by its eigenmodes that are determined by Eq.\,\ref{LAG:flux}, we can treat each dimer as a system of two oscillators with resonance frequencies $\omega_1$ and $\omega_2$ that are linearly coupled with rate $J_n$, similar to Ref.\,\cite{Eichler14_DPA}.
\begin{equation}
H = \hbar \omega_1 \mathbf{a}_1^\dagger \mathbf{a}_1 + \hbar \omega_2 \mathbf{a}_2^\dagger \mathbf{a}_2 + J_n \left( \mathbf{a}_1^\dagger \mathbf{a}_2 + \mathbf{a}_1 \mathbf{a}_2^\dagger  \right)
\end{equation} 
In this picture, the two oscillators correspond to the part of the array before and after the center capacitor, as seen from the input port. The measured eigenfrequencies $\omega_+$ and $\omega_-$ expressed in this new basis are
\begin{equation}
\omega_{\pm} = \frac{\omega_1 + \omega_2}{2} \pm \sqrt{\left( \frac{\omega_1 - \omega_2}{2} \right)^2 + J_n^2}.
\label{EQ_omegas}
\end{equation}
Since only the first part of the array is coupled to the input port with rate $\kappa$, a variation in the circuit parameters along the array ($\omega_1 \neq \omega_2$) will induce two different external coupling rates 
\begin{equation}
\kappa_\pm = \frac{\kappa}{2} \left(1 \pm \frac{\omega_1 - \omega_2}{\sqrt{4 J_n^2 + (\omega_1 - \omega_2)^2}} \right)
\label{EQ_kappas}
\end{equation}
with $\kappa = \kappa_+ + \kappa_-$. By inserting Eq.\,\ref{EQ_omegas} into Eq.\,\ref{EQ_kappas}, we derive an expression for the frequency asymmetry $A$ along the array, that depends only on the experimentally accessible quantities $\omega_\pm$ and $\kappa_\pm$
\begin{equation}
A = \left( \omega_1 - \omega_2 \right)^2 = \frac{(\kappa_+ - \kappa_-)^2 (\omega_+ - \omega_-)^2}{(\kappa_+ + \kappa_-)^2}.
\end{equation}
Finally, the bare frequencies $\omega_{1}$ and $\omega_2$ in the new basis are
\begin{equation}
\omega_{1,2} = \frac{\omega_+ + \omega_-}{2} \pm \frac{A}{2}.
\end{equation}

\section{Flux tunability}
\label{A_flux_tunability}
For a DJJAA device with $N = 1300$ SQUIDs, we investigated the flux tunability by measuring the power gain under different external bias flux conditions. Figure\,\ref{supfig_flux_tunability} a) depicts the phase response of the amplifier, similar to Fig.\,\ref{fig_gain_measurements} in the main text, where the main features are the second ($n = 2$), third ($n = 3$) and fourth ($n = 4$) dimer of the device. By changing the bias flux and adjusting the pump parameters (frequency and power), we observe signal power gain reaching $20\,\mathrm{dB}$ in a frequency range $\delta f \geq 1.5\,\mathrm{GHz}$ [cf. Fig.\,\ref{supfig_flux_tunability}b)] spanned between the highest and lowest lobes of the dimer for all three dimers.  

\begin{figure*}[t!]
\begin{center}
\includegraphics[width=6.67in]{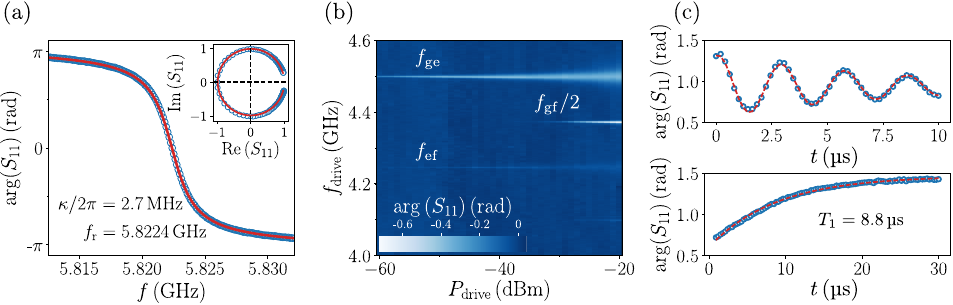}
\caption{\textbf{Transmon qubit characterization.} \textbf{a)} Phase of the complex reflection coefficient $\arg (S_{11})$ as a function of probe frequency $f$ for the readout resonator highlighted in Fig.\,\ref{fig_Transmon_sample}. The resonance frequency $f_\mathrm{r} = 5.8224\,\mathrm{GHz}$ and total linewidth $\kappa / 2\pi = 2.7 \, \mathrm{MHz}$ are extracted from a circle fit, which is indicated by the solid red line. As shown in the inset, the radius of the resonator response in the complex plane is approximately 1, implying $Q_\mathrm{i} \gg Q_\mathrm{c}$. \textbf{b)} Qubit spectrum measured by monitoring the resonator's phase response at resonance in a two-tone spectroscopy experiment. The first transition frequency appears at $f_\mathrm{ge} \approx 4.505 \, \mathrm{GHz}$ while the second transition frequency is $f_\mathrm{ef} \approx 4.244 \, \mathrm{GHz}$, resulting in an anharmonicity $\alpha_\mathrm{q} \approx 256\,\mathrm{MHz}$. Additionally, several multi photon transitions are visible. \textbf{c)} Rabi oscillations of the qubit (top panel) and energy relaxation measurement (bottom panel) following a $\pi-$pulse. The $\pi-$pulse duration of $1.5\,\si{\micro\second}$ correspondes to the first minimum of the measured Rabi oscillations. We extract an energy relaxation time of $T_1 = 8.8\,\si{\micro\second}$. From a Ramsey fringes measurement \cite{Ramsey50} [cf. Fig\,\ref{supfig_photon_calibration}], we extract a coherence time $T_2^\mathrm{\star} = 6.5\,\si{\micro\second}$.  }
\label{supfig_qubit_coherence}
\end{center}
\end{figure*}

\section{Qubit characterization}
\label{A_Qubit_chara}
For the coupled qubit-resonator system, the full Hamiltonian writes
\begin{equation}
\hat{H} = 4 E_\mathrm{c,q} \hat{N}^2 - E_\mathrm{J,q} \cos(\hat{\Theta}) + \hbar \omega_\mathrm{r} \hat{a}^\dagger \hat{a} + \hbar g \hat{N}\left( \hat{a} + \hat{a}^\dagger \right), 
\label{Ham:Cooper}
\end{equation}
where $\hat{N}$ is the number operator of the Cooper Pairs (CP) transferred over the Josephson junction, $\hat{\Theta}$ is the superconducting phase difference across the JJ, $E_\mathrm{c,q}$ is the charging energy due to its shunt capacitance and $E_\mathrm{J,q}$ is the Josephson energy. Furthermore, $\hat{a}^\dagger$ and $\hat{a}$ are the creation and annihilation operators of the bare cavity mode with angular frequency $\omega_\mathrm{r}$. The electric field of the resonator $\vec{E} \propto (\hat{a}^\dagger + \hat{a})$ couples to the qubit's electric dipole moment with a coupling rate $g$. \\
For a ratio $E_\mathrm{J,q} / E_\mathrm{c,q} \geq 50$, usually referred to as the transmon regime, the qubit itself is described by a non-linear resonator with bare resonance frequency $\omega_\mathrm{q}$ and relatively small anharmonicity $\alpha_\mathrm{q} \ll \omega_\mathrm{q}$. In the dispersive limit, the qubit is weakly coupled to the resonator $g \ll \Delta = \vert \omega_\mathrm{r} - \omega_\mathrm{q} \vert$, only imposing a small frequency shift $\chi_\mathrm{qr}$ on the resonator according to its state. Hence, for small probe powers and weak coupling, the system Hamiltonian given in Eq.\,\ref{Ham:Cooper} is simplified by considering a reduced Hilbert-space only \cite{Nigg12,Lescanne18}, 
\begin{equation}
\begin{split}
\hat{H}_\mathrm{low} / \hbar \,\, = \,\, &\tilde{\omega}_\mathrm{q} \hat{a}_\mathrm{q}^\dagger \hat{a}_\mathrm{q} - \frac{\alpha_\mathrm{q}}{2}\left( \hat{a}_\mathrm{q}^\dagger \right)^2 \hat{a}_\mathrm{q}^2 - \chi_\mathrm{qr} \hat{a}_\mathrm{q}^\dagger \hat{a}_\mathrm{q} \hat{a}_\mathrm{r}^\dagger \hat{a}_\mathrm{r} \\
& + \tilde{\omega}_\mathrm{r} \hat{a}_\mathrm{r}^\dagger \hat{a}_\mathrm{r} - \frac{\alpha_\mathrm{r}}{2} \left( \hat{a}_\mathrm{r}^\dagger \right)^2 \hat{a}_\mathrm{r}^2,	 
\end{split}
\label{Ham:Cooper_BBQ}
\end{equation}
where $\hat{a}^\dagger_\mathrm{i}$ and $\hat{a}_\mathrm{i}$ with $i = \left( \mathrm{q}, \mathrm{r} \right)$ are the creation and annihilation operator of the dressed fundamental qubit and readout resonator modes, with angular frequencies $\tilde{\omega}_\mathrm{q}$ and $\tilde{\omega}_\mathrm{r}$ and self-Kerr coefficients $\alpha_\mathrm{q}$ and $\alpha_\mathrm{r}$, respectively. The cross-Kerr between both modes, denoted $\chi_\mathrm{qr}$, determines the dispersive shift of the resonator induced by the qubit state and the dependence of the qubit's first transition frequency $\tilde{\omega}_\mathrm{q}$ on the mean number of photons circulating in the resonator $\bar{n}_\mathrm{r}$. 
\begin{equation}
\tilde{\omega}_\mathrm{q} ( \bar{n} ) =  \tilde{\omega}_\mathrm{q}^\mathrm{0} - \chi_\mathrm{qr} \bar{n}_\mathrm{r}
\label{EQ:ACStark}
\end{equation}
From a circle fit to the complex response of the readout resonator, see Fig.\,\ref{supfig_qubit_coherence}a), we extract the resonator frequency $\tilde{\omega}_\mathrm{r} / 2 \pi = 5.8224\,\mathrm{GHz}$ and total linewidth $\kappa_\mathrm{r} / 2 \pi \approx 2.7 \,\mathrm{MHz}$. By applying a second tone at frequency $f_\mathrm{drive}$ to the qubit while continuously monitoring the reflected phase of the readout resonator at resonance, we measure the qubit spectrum [cf. Fig.\,\ref{supfig_qubit_coherence}b)]. The first transition frequency $f_\mathrm{ge} = \tilde{\omega}_\mathrm{q} / 2 \pi$ is found at $f_\mathrm{drive} = 4.505\,\mathrm{GHz}$, with an anharmonicity $\alpha_\mathrm{q} / 2 \pi = 256\,\mathrm{MHz}$ compared to the next higher transition $f_\mathrm{ef}$. For higher drive powers $P_\mathrm{drive}$, several multi-photon transitions appear. The dispersive shift is found to be $\chi_\mathrm{qr} / 2 \pi = 480 \, \mathrm{kHz}$ from a histogram of the $IQ$-plane, by separating the ground and first excited state of the qubit with the aid of our DJJAA [cf. Fig.\,\ref{fig_Quantum_jumps} in the main text]. \\
\\
\begin{figure*}[t!]
\begin{center}
\includegraphics[width=6.67in]{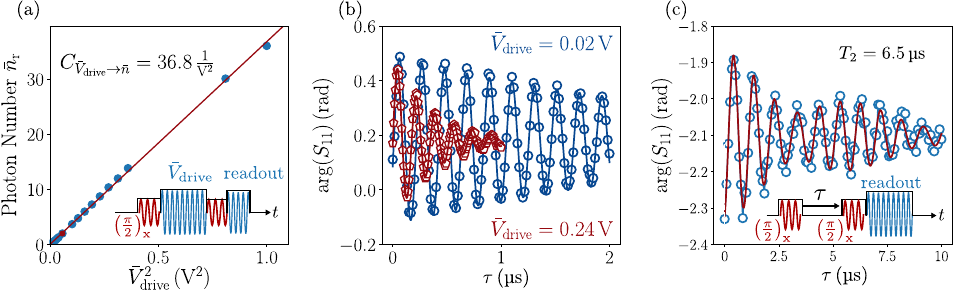}
\caption{\textbf{a)} Photon number calibration of the readout strength $\bar{V}_\mathrm{drive}^2$. The pulse sequence used in the experiment is depicted in the lower right corner. In-between two $(\pi / 2)_x$ pulses applied to the qubit, we apply an additional tone to the readout resonator with varying voltage amplitude $\bar{V}_\mathrm{drive}$, before we readout the resonator (qubit state). From the change in frequency $f_\mathrm{R}$ observed in the Ramsey fringes [cf. panel b)] with respect to the undriven case $f_\mathrm{R}^0 = |f_\mathrm{q, drive} - f_\mathrm{q}|$, and using the dispersive shift $\chi_\mathrm{qr}$ [cf. Fig.\,\ref{fig_Quantum_jumps}], we calibrate the photon number $\bar{n}_\mathrm{r}$ as a function of the applied drive power $P_\mathrm{drive} \propto \bar{V}_\mathrm{drive}^2$. \textbf{b)} Two examples of measured Ramsey fringes with $\bar{V}_\mathrm{drive} = 0.02\,\mathrm{V}$ (blue) and $\bar{V}_\mathrm{drive} = 0.24\,\mathrm{V}$ (red). Besides the change in the frequency $f_\mathrm{R}$, the measured coherence time $T_2^\mathrm{\star}$ decreases also due to measurement induced dephasing. \textbf{c)} Standard Ramsey fringes experiment without an additional drive applied to the readout resonator. Notice the longer time scale compared to panel b); the pulse sequence is depicted in the lower right hand corner. We observe two oscillations with similar frequency $f_\mathrm{R}^1 = 1.03\,\mathrm{MHz}$ and $f_\mathrm{R}^2 = 1.19\,\mathrm{MHz}$, with a characteristic decay time $T_2^\mathrm{\star} = 6.5\,\si{\micro\second}$.}
\label{supfig_photon_calibration}
\end{center}
\end{figure*}
In order to extract the circuit parameters $E_\mathrm{J,q}$, $E_\mathrm{c,q}$, $g$ and $\omega_\mathrm{r}$ of the system Hamiltonian given in Eq.\,\ref{Ham:Cooper}, we numerically solve for the lowest eigenenergies in the charge basis for the qubit and the Fock basis for the readout resonator $\ket{N,n}$ \cite{Lescanne18}. We compare the results to the experimentally observed values measured at low readout powers, where the system is approximated by the Hamiltonian in Eq.\,\ref{Ham:Cooper_BBQ}. We find $E_\mathrm{J,q} / h = 12.5\,\mathrm{GHz}$, $E_\mathrm{c,q} / h = 225\,\mathrm{MHz}$ $\left( E_\mathrm{J,q} / E_\mathrm{c,q} \approx 56 \right)$ and $g  = 39 \,\mathrm{MHz}$, which are in good agreement with values extracted from finite element simulations for the charging energy and room temperature resistance measurements for the Josephson energy. \\
\\
By applying a drive tone to the qubit at its fundamental transition frequency $\tilde{f}_\mathrm{q}$ with rectangular envelope of varying duration $t$, we observe continous Rabi-oscillations between the qubit's ground and first excited state [cf. top panel Fig.\,\ref{supfig_qubit_coherence}c)]. Using relatively weak drive strength, we calibrate the duration of a $\pi$-pulse to $t_\pi \approx 1.43\,\si{\micro\second}$ by fitting the data (blue) with a periodic cosine function with exponentially decaying envelope (red). By exciting the qubit to its first excited state and gradually increasing the time interval $t$ between the end of the excitation pulse and the beginnig of the readout pulse, we perform an energy relaxation measurement. From an exponential fit (red) to the data (blue), the corresponding energy relaxation time is found to be $T_1 \approx 8.8\,\si{\micro\second}$ [cf. bottom panel Fig.\,\ref{supfig_qubit_coherence}c)]. The qubit's coherence time, extracted from a Ramsey fringes measurement, is $T_2^\mathrm{\star} \approx 6.5\,\si{\micro\second}$ [cf. Fig.\,\ref{supfig_photon_calibration}].

\section{Photon number calibration}
\label{A_photon_number}
Figure\,\ref{supfig_photon_calibration} a) depicts the calibrated photon number as a function of the input power. As expected, the photon number increases linearly with the drive power $P_\mathrm{drive} \propto \bar{V}_\mathrm{drive}^2$. The change in the Ramsey frequency $f_\mathrm{R}$ and the observed coherence time $T_2^\mathrm{\star}$ is illustrated in Fig.\,\ref{supfig_photon_calibration}b) for two different drive strengths. Notably, the decrease in coherence results from an increasing distribution of the coherent state in the Fock-basis, which is proportional to $\bar{V}_\mathrm{drive}$ \cite{Gambetta06}. We fit the measured oscillations with an exponentially damped cosine
\begin{equation}
\varphi ( \tau) = e^{-\tau / T_2^\mathrm{\star}} A \cos \left[2 \pi (f_\mathrm{R} \tau) + \phi_0 \right] + \varphi_0.
\end{equation}
Here $\tau$ is the evolution time during the experiment, $T_2^\mathrm{\star}$ is the coherence time, $A$ is the oscillation amplitude, $f_\mathrm{R}$ is the Ramsey frequency, $\phi_0$ is the offset phase of the cosine oscillation and $\varphi_0$ is the global offset phase. In the case of two distinct Ramsey frequencies, as depicted in Fig.\,\ref{supfig_photon_calibration}, we use an extended fit function $\varphi ( \tau)$	
\begin{align}
\begin{split}
\varphi ( \tau) = & \exp^{-\tau / T_2^\mathrm{\star}} (  A_1 \cos \left[2 \pi (f_\mathrm{R,1} \tau) + \phi_{0,1} \right]  \\
& + A_2 \cos \left[2 \pi (f_\mathrm{R,2} \tau) + \phi_{0,2} \right] ) \varphi_0. 
\end{split}
\end{align}
Most probably, the two distinct Ramsey frequencies $f_\mathrm{R,1}$ and $f_\mathrm{R,2}$ arise from two distinct qubit frequencies. However, the origin of these two distinct qubit frequencies remains unknown.

\section{Qubit temperature}
\label{A_qubit_temperature}
The effective temperature of the qubit $T_\mathrm{q}$ [cf. Fig.\,\ref{fig_Quantum_jumps}] is calculated from the relative occupation of the qubit states with eigenenergy $E_k$. For a superconducting charge qubit in the transmon regime ($E_\mathrm{J,q} / E_\mathrm{c,q} \geq 50$), the eigenenergies are approximately given by \cite{Koch07}
\begin{equation}
E_k \approx -E_\mathrm{J,q} + \sqrt{8 E_\mathrm{J,q} E_\mathrm{c,q}} (k + \frac{1}{2}) - \frac{E_\mathrm{c,q}}{12} (6 k^2 + 6 k + 3 ).
\end{equation}
The deviations from numerical solutions of the exact charge qubit Hamiltonian, given in Eq.\,\ref{Ham:Cooper}, are found to be neglectable for qubit states inside the Josephson potential. For a thermal state, the occupation $N_k$ of the $k$-th energy state follows a Boltzmann distribution
\begin{equation}
N_k \propto e^{-\frac{E_k}{k_\mathrm{B} T_\mathrm{q}}},
\end{equation}
where $k_\mathrm{B}$ is the Boltzmann constant. Under this assumption, we calculate the effective qubit temperature from the Boltzmann factor of the first two eigenstates
\begin{equation}
T_\mathrm{q} = \frac{E_1 - E_0}{k_\mathrm{B} \ln (N_0 / N_1)},
\end{equation}
if we cannot distinguish more than two qubit states, or by fitting a Boltzmann distribution to the relative occupation
\begin{equation}
N_k (\epsilon^\prime) / N = n_0 e^{-\frac{\epsilon_k}{k_\mathrm{B} T_\mathrm{q}}},
\end{equation}
with the parameter $\epsilon_k = (E_k - E_0)$. \\

\end{document}